\documentclass{aa}
\usepackage[varg]{txfonts}
\usepackage{tablefootnote}
\usepackage{color}

\begin{document}
\title{Following the TraCS of exoplanets with Pan-Planets: Wendelstein-1b and Wendelstein-2b 
\thanks{Based on observations obtained with the Hobby-Eberly Telescope, which is a joint project of the University of Texas at Austin, the Pennsylvania State University, Stanford University, Ludwig-Maximilians-Universität München, and Georg-August-Universität Göttingen.}
}

\author{C. Obermeier\inst{1,2}, J. Steuer \inst{1,2}, H. Kellermann \inst{2,1}, R. P. Saglia\inst{2,1}, Th. Henning \inst{3}, A. Riffeser \inst{1}, U. Hopp \inst{1,2}, G. Stefansson\inst{4,5,6,7,8}, C. Ca\~nas\inst{4,5,6,9}, J. Ninan\inst{4}, S. Mahadevan\inst{4,5}, H. Isaacson\inst{10}, A. W. Howard\inst{11}, J. Livingston\inst{11,12}, J. Koppenhoefer\inst{2,1}, R. Bender\inst{1,2}}
\institute{{University Observatory Munich (USM), Scheinerstraße 1, D-81679 Munich, Germany }
\and
{Max-Planck-Institute for Extraterrestrial Physics, Garching, Gießenbachstraße, D-85741 Garching, Germany }
\and
{Max-Planck-Institute for Astronomy, Heidelberg, Königstuhl 17, D-69117 Heidelberg, Germany }
\and 
Department of Astronomy \& Astrophysics, The Pennsylvania State University, 525 Davey Lab, University Park, PA 16802, USA
\and Center for Exoplanets \& Habitable Worlds, University Park, PA 16802, USA
\and NASA Earth and Space Science Fellow
\and Department of Astrophysical Sciences, Princeton University, 4 Ivy Lane, Princeton, NJ 08540, USA
\and Henry Norris Russell Fellow
\and Penn State Astrobiology Research Center, University Park, PA 16802, USA
\and Department of Astronomy, University of California, Berkeley CA 94720, USA
\and California Institute of Technology, Pasadena, CA 91125, USA
\and Department of Astronomy, The University of Tokyo, Hongo 7-3-1, Bunkyo-ku, Tokyo, 113-0033, Japan
}

\abstract{Hot Jupiters seem to get rarer with decreasing stellar mass. The goal of the Pan-Planets transit survey was the detection of such planets and a statistical characterization of their frequency. Here, we announce the discovery and validation of two planets found in that survey, Wendelstein-1b and Wendelstein-2b, which are two short-period hot Jupiters that orbit late K host stars. We validated them both by the traditional method of radial velocity measurements with the HIgh Resolution Echelle Spectrometer (HIRES) and the Habitable-zone Planet Finder (HPF) instruments and then by their Transit Color Signature (TraCS). We observed the targets in the wavelength range of $4000\AA - 24000\AA $ and performed a simultaneous multiband transit fit and additionally determined their thermal emission via secondary eclipse observations. Wendelstein-1b is a hot Jupiter with a radius of $1.0314_{-0.0061}^{+0.0061}$\,$R_J$ and mass of $0.592_{-0.129}^{+0.165}$\,$M_J$, orbiting a K7V dwarf star at a period of 2.66\,d, and has an estimated surface temperature of about $1727_{-90}^{+78}$\,K. Wendelstein-2b is a hot Jupiter with a radius of $1.1592_{-0.0210}^{+0.0204}$\,$R_J$ and a mass of $0.731_{-0.311}^{+0.541}$\,$M_J$, orbiting a K6V dwarf star at a period of 1.75\,d, and has an estimated surface temperature of about $1852_{-140}^{+120}$\,K. With this, we demonstrate that multiband photometry is an effective way of validating transiting exoplanets, in particular for fainter targets since radial velocity (RV) follow-up becomes more and more costly for those targets.}

\date{Version: May 27, 2020}
\authorrunning{C. Obermeier et al.}
\titlerunning{Wendelstein-1b and Wendelstein-2b}
\maketitle
\section{Introduction}
The field of exoplanets is undergoing rapid changes with the impact of space-based telescopes and their sheer number of candidate discoveries. 
While \textit{Kepler} \citep{2010Sci...327..977B} and K2 \citep{2014PASP..126..398H} alone provide several thousands of confirmed planets and even more candidates, each of those candidates had to be verified in a time-consuming and costly process.

The percentage of genuine exoplanets that are confirmed by transit surveys greatly depends on the particular survey.
For the \textit{CoRoT} mission (\cite{baglin2006}), for example, the false positive rate was found to be close to $90\%$, considering candidates whose nature was resolved. This rate is composed of on target eclipsing binaries (EB), which make up $53.7\%$ of the observed targets, background EB systems, which dilute the target system and account for $36.1\%$ and the $10.2\%$ of the candidates which were found to be brown dwarfs (\cite{deleuil2018}). The false positive probability (FPP) for the \textit{Kepler} mission has been discussed at great lengths. Initially, \cite{morton2011} estimated a FPP of below $10\%$ for $90\%$ of the Kepler objects of interest (KOI) and half of those had a FPP below $5\%$. Observations indicated a higher rate, such as \cite{santerne2012}, who found a FPP of $34.6 \% \pm 6.5 \%$ for giant close-in \textit{Kepler} candidates that were observed with the SOPHIE spectrograph at the Observatoire de Haute-Provence. Later, the overall FPP for Kepler was re-estimated by \cite{fressin2013}, who modeled the efficiency of the \textit{Kepler} pipeline to detect planets. They find a global FPP of $9.4 \% \pm 0.9 \%$, which peaks for giant planets at $17.7\%$ and reaches a minimum for small Neptunes (2 to 4 Earth radii) at $6.7\%$. This value was later re-evaluated by \cite{santeren2013}, who find a global FPP of $11.3 \% \pm 1.1 \%$ due to eclipsing binaries for which only the secondary event could be observed from Earth.
This offset of FPP for the \textit{CoRoT} and \textit{Kepler} missions can mainly be attributed to the ability of the \textit{Kepler} pipeline to precisely measure centroid offsets and thereby quickly discard background eclipsing binaries and background planetary transits before they have to be followed up on. 

There are several ways a false positive signal can be created, even more so when observing with very broad optical filters as in the case of \textit{Kepler} and \textit{TESS} \citep{2015JATIS...1a4003R}.
Further confounding this problem is the fact that many of the most intriguing candidates are either very faint or show very shallow eclipses and radial velocity amplitudes. This leads to difficulties for a radial velocity (RV) follow-up, which for transiting signals is the classical path toward the confirmation of the planet status. An RV curve, when measured to sufficient precision, grants insights about the stellar companion's mass and orbit characteristics, which can be a great tool in determining the overall nature of the planet system when the two methods are combined. 
However, large 8m-class telescopes and upward are required for the fainter planet candidates and even they may not suffice, as was the case for some of the planet candidates discovered in the Pan-Planets survey \citep{2016A&A...587A..49O}. 

Time on 1-2m class telescopes is abundant and cheap in comparison to larger-class telescopes and this time can be spent as an effective way to rule out all false positive scenarios, something that is particularly useful for managing the massive data output of the current surveys.
We therefore chose to facilitate the confirmation process by supplementing RV measurements with multiband transit observations using 3KK \citep{2016SPIE.9908E..44L}, a 3-channel camera mounted on the 2.1m Fraunhofer Telescope Wendelstein (FTW, \cite{2010SPIE.7733E..07H}) at the Wendelstein observatory in the Bavarian Alps.

Evidence for a false positive detection can be found when comparing the chromatic transit depth at different wavelengths \citep{article} due to the characteristics of limb darkening, resulting in different eclipse depths or shapes, depending on the false positive scenario. Each alternative false positive scenario has its own set of constraints for inclination, radius ratio and presence of blended background sources. Thanks to multiband observations of the primary and secondary eclipses, those parameters can be disentangled and the true scenario be determined. 
However, it is to be noted that even for a true planetary detection, the measurement of some parameters, for example the planetary radius and therefore the transit depth, can still be wavelength dependent due to atmospheric scatter or absorption. These effects are nevertheless relatively small compared to the varying impact of a false positive scenario like an eclipsing binary on the different wavebands. 

Multiband photometry gives the additional advantage of assessing the stellar parameters more precisely than singleband by attaining a combined fit for the orbital parameters. We modify this commonly known concept by using near-infrared (NIR) bands in our observations, which allows us to observe the secondary eclipse as well and measure the planet's effective temperature or at least place upper limits on them. Furthermore, we utilize 3KK's relatively high resolution of 0.2/0.24 arcsec/px for the optical and NIR bands, respectively, to perform a multiband centroid analysis, giving much more strict limits on possible background source contamination. 

There is a population of gaseous giants that appears to be quite rare: hot Jupiters in cool ($T_{\rm eff} < 4500$ K) systems, of which only a few are known such as Kepler-45b \citep{2012AJ....143..111J}, WASP-80b \citep{2013A&A...551A..80T}, or HATS-6b \citep{2015AJ....149..166H}. In the context of our ongoing follow-up survey, we announce the discovery of two new planets, which were both confirmed by the traditional way of radial velocity observations and by photometric multiband observations of primary plus secondary eclipses, combined with centroid analyses in all bands. In Sect. 2, we provide an overview of the previous candidate vetting and follow-up efforts for Wendelstein-1b and Wendelstein-2b. The photometric follow-up observations with the Wendelstein 3-channel imager 3KK and our data processing is described in Sect. 3, Sect. 4 details the spectroscopic pendant. We verify our planets in Sect. 5, perform a multiband fit and discuss potential contamination by blends and conclude with a discussion in Sect. 6.
\section{Candidate vetting}
\subsection{Pan-Planets}
Wendelstein-1b and Wendelstein-2b were part of the exoplanet candidate set of Pan-Planets, a survey that was performed by Pan-STARRS1 (\cite{2004AN....325..636H}, \cite{2016arXiv161205560C}) (PS1), a wide-field, high-resolution telescope located at the Haleakala Observatory on Maui, Hawaii. The survey was granted 4\% of the overall observing time of PS1. Starting in May 2010, the survey, undertaken with an effective field of view (FoV) of 42 square degrees, pursued the goal of finding transiting planets, which orbit main sequence stars with a special focus on the coolest stellar types of M and K. Analysis of the sensitivity of the survey was conducted by utilizing Monte Carlo simulations (\cite{2009A&A...494..707K}) and yielded a prognosis of up to dozens of Jupiter-sized planets being detectable, depending on observing strategies and photometric stability of the telescope. 

However, the larger than expected amount of red noise residuals, which capped the photometric quality at about 5mmag instead of the expected 2mmag, meant that the number of detectable targets was significantly decreased. 
This increase in noise can be generally traced back to bad processing and worse system performance than advertised. One concrete aspect was the sub-optimal flat fielding between the individual CCD cells, which led to inconsistent backgrounds. In combination with permanent wide dithering, the initially predicted efficiency of 1-3mmag turned into around 10 mmag unclipped accuracy after a sizable effort to clean up the data. Additionally, the fields of observations for the Pan-Planets survey were selected very close to the galactic disk, which means on one hand, more stars are present in the FoV, however, at the same time also more dust, which contributes to the amount of red noise. The acquired data were reduced by the PS1 Image Processing Pipeline (IPP, \cite{2006amos.confE..50M}) on site in Hawaii, where all images were processed in their standard reduction steps such as de-biasing, flat-fielding, astrometric calibration and pixel to WCS transformation . Later, the data were transferred and stored in Germany, undergoing analysis within the Astronomical Wide-field Imaging System for Europe (Astro-WISE) environment (\cite{2013ExA....35....1B}), where the final data processing steps such as a correction of systematics and time-series photometry were taken as well. For a more detailed recount of this process, we refer to \citet{2016A&A...587A..49O}. 

The images, varying between 1500-4500 per pointing due to overlapping regions, were analyzed using the difference imaging method. The resulting photometric data for each star were then subsequently searched for periodic signals with an algorithm based on box-fitting least squares (BLS) of \cite{2002A&A...391..369K}, which we further modified to additionally fit secondary eclipses and trapezoids instead of simple boxes to both \citep{2013A&A...560A..92Z}. This yielded a number of high-priority candidates to be be followed up. In this work, we focus on the two primary candidates that are of scientific interest for their rare combination of host star and planet type. 

Having condensed the candidate set to four high-priority targets, we performed the final stage of confirmation: first, conducting a multiband survey and then radial velocity follow-up observations for the successfully confirmed candidates. Early on in the 3KK survey, we had to disregard two of these candidates since they showed clear signs of being false-positive EB detections due to their eclipse depths and shapes of multiband fits which were not in agreement with the planet scenario. In the following, we therefore focus on the successful planet detections, Wendelstein-1b and Wendelstein-2b. 

\subsection{Wendelstein WWFI survey}
From April to September 2015 the selected candidates from the initial Pan-Planets survey were followed up with the Wendelstein Wide Field Imager (WWFI, \cite{2014ExA....38..213K}) on the 2.1m FTW. WWFI is a 64 megapixel camera with a FOV of 0.5 deg$^2$ , divided onto $2x2$ CCDs which each consist of $2x2$ cells with a resolution of 4 megapixels each and a pixel scale of 0.2 arcsec/pixel. For the observations, the i'-band was chosen as it is very similar to the band used in the original Pan-Planets observations. After the data collection was complete, the custom Wendelstein data reduction pipeline was utilized to automatically perform basic reduction steps like flat fielding, saturation masking, sky subtraction, charge persistence corrections, dark and bias processing, astrometric identification and regridding. Aperture photometry with optimized parameters for each night yielded light curves for the candidates selected by the Pan-Planets survey. The bottom frames of Fig. \ref{fig:pp140-14711} and \ref{fig:pp127-21645} show the light curves yielded by the Pan-Planets survey (black dots) and the initial WWFI follow-up (red diamonds) for the candidates PP140-14711/Wendelstein-1b and PP127-21645/Wendelstein-2b, respectively, including the best-fit model (blue line). Based on this survey and its improved photometry over Pan-Planets, we already disregarded a number of candidates as false positive detections of eclipsing binaries.

\subsection{Stellar classification}
We utilized the spectral energy distribution (SED) fitting code from \citet{2016A&A...587A..49O} as control for our stellar type characterization. In contrast to spectroscopy, this approach relies on broad-band photometry. We extracted the Pan-STARRS1 DR2 data for our targets and cross-matched their coordinates with the 2MASS catalog. For the synthetic stellar SED catalog, we used the newest version of the PARSEC isochrones package \citep{2012MNRAS.427..127B}. As described in \citet{2016A&A...587A..49O}, we combined and interpolated those isochrones with the Dartmouth \citep{2008ApJS..178...89D}, BT-Dusty \citep{2012RSPTA.370.2765A} and \cite{2015A&A...577A..42B} isochrones. Furthermore, we made use of the distance-dependent extinction values given in the 3D dust map from \cite{2015ApJ...810...25G}, which has recently been modified to use Gaia data as well \citep{2019ApJ...887...93G}. We interpolated with a 10th-order polynomial and performed an iterative fit for distance and extinction until both converged. The reason for the interpolation is the fact the extracted dust map contains about 90 extinction points per star and linear interpolation led to artifacts in tests. Using a polynomial was the preferred choice due to numerical considerations. The entire survey encompassed millions of stars and each was fit individually, and 10th order was determined to be the best compromise between accuracy and the introduction of higher order systematics intrinsic to polynomials. A comparison with the updated, Gaia-assisted version yielded no different result for our targets.

For spectroscopic follow-up and validation of the SED fitting technique, we utilized the low-resolution Cassegrain Spectrometer (ES2) \citep{2014acm..conf..446R}, mounted on the 2.1m Otto-Struve telescope on the McDonald Observatory on Mt. Locke, between July 14 and 24 2015. Since the target stars are very faint, we observed with lower resolution gratings at 600 grooves per millimeter. The spectra were processed with a custom in-house data reduction pipeline, wavelength-calibrated with a Neon-Argon lamp, and then fit against stellar templates that we created with the SYNSPEC IDL interface \citep{1995ApJ...439..875H}, convolved into the resolution of the spectrograph. 
We furthermore queried the Gaia DR2 catalog \citep{2018A&A...616A...1G} for accurate parallaxes and their own stellar type characterization. Furthermore, we searched the catalog for the presence of any nearby sources that may contaminate our photometry, which is discussed in more detail in Sect. \ref{sec:pot_contamination}.

\begin{figure}
  \centering
  \includegraphics[width=0.9\linewidth]{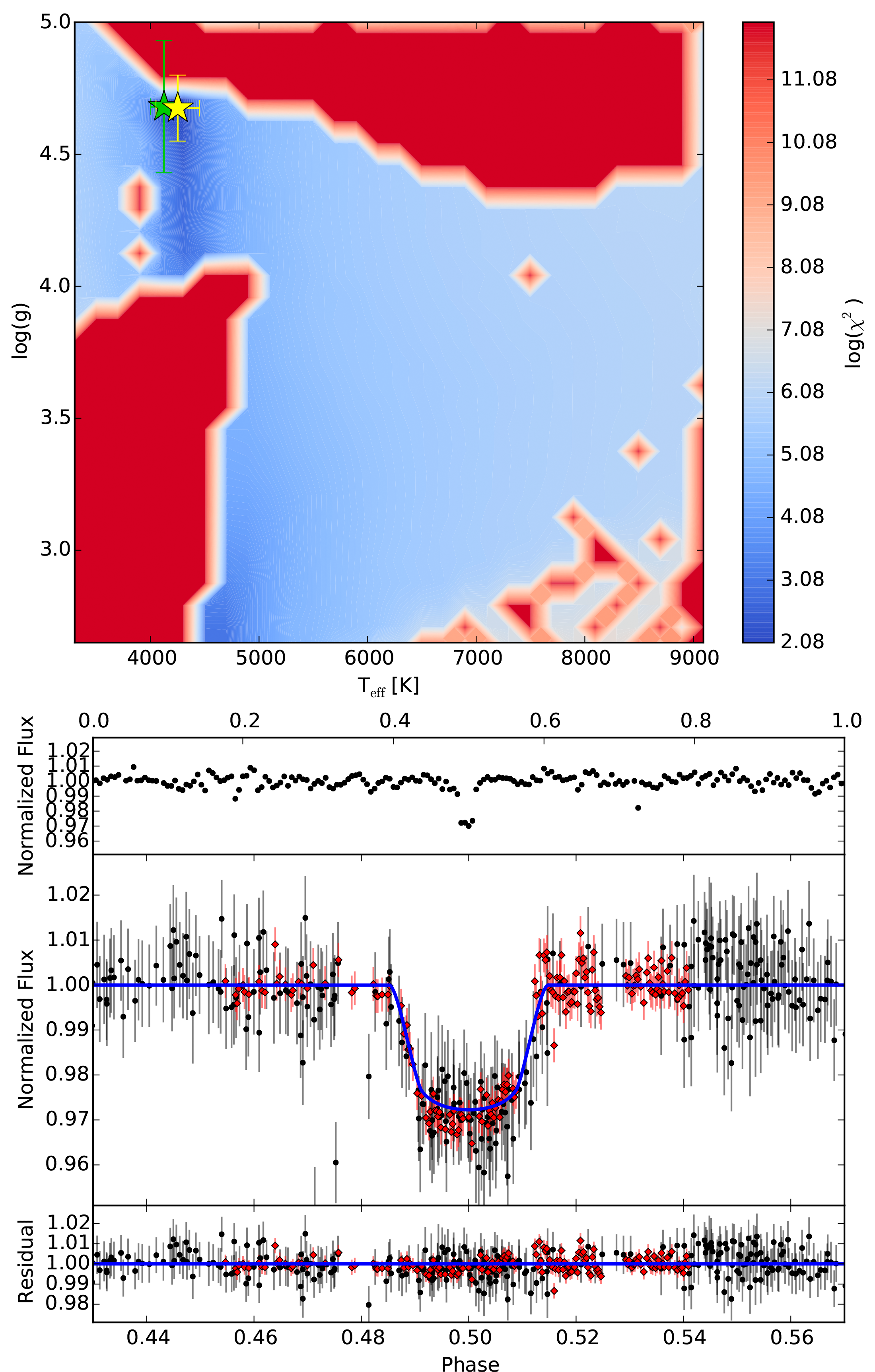}
  \caption{ Automatically created post-vetting data for Wendelstein-1b. Top: stellar parameter fits with the results from SED fitting (yellow) and spectroscopy (green) with an underlying $\chi^2$ contour plot of log g against effective temperature based on SED fitting. Bottom: folded light curve with the best-fitting model (blue line) and data from Pan-Planets (black circles) and WWFI of the 2m Wendelstein telescope (red diamonds). Note: the internal convention of Pan-Planets was to place the primary transit at phase 0.5; the y-axis of the residual frame was by default zoomed out to better assess the goodness of the fit.}
  \label{fig:pp140-14711}
\end{figure}

\begin{figure}
  \centering
  \includegraphics[width=0.9\linewidth]{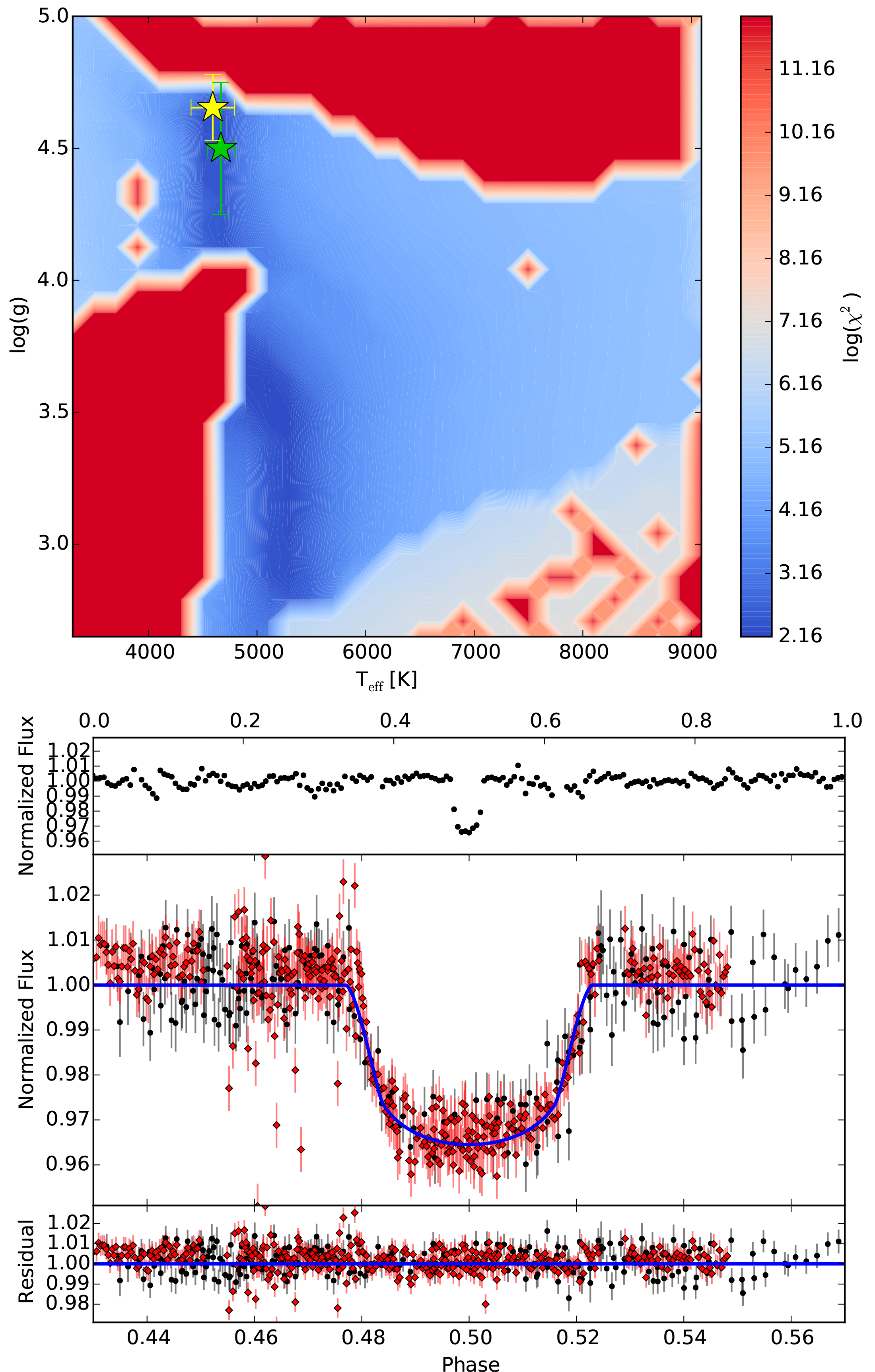}
  \caption{ Automatically created post-vetting data for Wendelstein-2b. Top: stellar parameter fits with the results from SED fitting (yellow) and spectroscopy (green) with an underlying $\chi^2$ contour plot of log g against effective temperature based on SED fitting. Bottom: folded light curve with the best-fitting model (blue line) and data from Pan-Planets (black circles) and WWFI of the 2m Wendelstein telescope (red diamonds). Note: the internal convention of Pan-Planets was to place the primary transit at phase 0.5; the y-axis of the residual frame was by default zoomed out to better assess the goodness of the fit.}
  \label{fig:pp127-21645}
\end{figure}

The results of this follow-up study are shown in the top frames of Fig. \ref{fig:pp127-21645} and Fig. \ref{fig:pp140-14711}, respectively, where the results from SED fitting (yellow) and from spectroscopy (green) agree well on the stellar type classification. Further information about the host stars can be gathered from Table \ref{tab:pp140-14711-stats}. The effective temperature measurements based on Gaia are systematically lower than those based on our own measurements. A possible explanation for this may be extinction that is being accounted for differently, since the effect is more pronounced for the higher-extinct Wendelstein-2. For Wendelstein-1 a temperature of 4251\,K and radius of 0.61 $\pm$ 0.09\,R$_\odot$ and for Wendelstein-2, a temperature of 4592\,K and radius of 0.66 $\pm$ 0.10\,R$_\odot$ were determined, respectively. According to these measurements, they can therefore both be categorized as late K-dwarfs. However, metallicity was difficult to constrain due to the low S/N of the spectra.
\begin{table}[hbt]
  \centering
  \begin{tabular}{c|c|c|c} 
  & Wendelstein-1 & Wendelstein-2 & \\
  \hline
   \textit{ID}   &  PP140\_14711 & PP127\_21645 &\\
   \textit{RA}  & 19h56m12.15s & 19h51m09.81s &1\\
   \textit{DEC} & 17d34m11.86s & 16d54m45.17s &1\\
   \textit{Epoch \& Equinox} & J2000.0 & J2000.0 &1\\ 
   \textit{T$_{spec}$} [K] & 4125 $\pm$ 210 & 4667 $\pm$ 238& 2 \\
   \textit{T$_{phot}$} [K]& 4251 $\pm$ 175 & 4591 $\pm$ 195& 3 \\
   \textit{T$_{Gaia}$} [K]& 3984$_{-46}^{+152}$ & 4288$_{-111}^{+133}$ & 4 \\
   \textit{ $\pi$}* [mas] & 3.247 $\pm$ 0.032 & 1.741 $\pm$ 0.043 & 4 \\
   \textit{d} [pc] & 304 $\pm$ 4 & 575 $\pm$ 6 & 4\\
   \textit{E(g-r)} & 0.06$_{-0.02}^{+0.03}$ & 0.12$_{-0.01}^{+0.04}$ & 5\\
   \textit{R} [$R_{\odot}$]  & 0.61 $\pm$ 0.09 & 0.66 $\pm$ 0.10& 3 \\
   \textit{M} [$M_\odot$] & 0.65 $\pm$ 0.10 & 0.73 $\pm$ 0.11& 3 \\
   \textit{log g} & 4.68 $\pm$ 0.12 &4.66 $\pm$ 0.13 & 3 \\
   \textit{g'} [mag] & 16.232$\pm$ 0.005 & 16.97 $\pm$ 0.016& 1 \\
   \textit{r'} [mag] & 15.093 $\pm$	0.003 & 15.87 $\pm$ 0.007& 1 \\
   \textit{i'} [mag] & 14.519 $\pm$	0.002 & 15.43 $\pm$ 0.012& 1 \\
   \textit{z'} [mag] & 	14.262 $\pm$	0.004 & 15.20 $\pm$ 0.014 & 1 \\
   \textit{J} [mag] & 12.971 $\pm$ 0.018 & 14.06 $\pm$ 0.029 & 6 \\
   \textit{H} [mag] & 12.331 $\pm$ 0.018 & 13.43 $\pm$ 0.029 & 6\\
   \textit{Ks} [mag] & 12.198 $\pm$ 0.020 & 13.33$\pm$ 0.039 & 6\\
  \end{tabular}
  \caption{Stellar properties of Wendelstein-1 and Wendelstein-2 based on the initial candidate vetting process. References are: 1 - PS1 catalog \citep{2016arXiv161205243F}; 2 - ES2 spectroscopic data; 3 - SED fitting; 4 - GAIA DR2 catalog \citep{2018A&A...616A...1G}; 5 - 3D dust map \citep{2015ApJ...810...25G}; 6 - 2MASS catalog \citep{2006AJ....131.1163S}. * parallax. Note: the uncertainties for \textit{T$_{spec}$} show 2$\sigma$ values since the fitting routine seemed to have underestimated the noise due to the low spectral resolution.}
  \label{tab:pp140-14711-stats}
\end{table}
\section{Follow-up observations}
\subsection{3KK survey}
3KK \citep{2016SPIE.9908E..44L} is equipped with two Apogee ALTA F3041 cameras for optical and one H2RG CMOS for NIR light, therefore allowing to make simultaneous observations in three different wavebands. With a FoV of 7 $\times$ 7 arcmin for the optical and 8 $\times$ 8 arcmin for the NIR filters, the pixel scale for the optical filters is approximately 0.2 arcsec/pixel and for the NIR filters 0.24 arcsec/pixel. For the blue CCD, the Sloan-like filters u', g' and r' can be deployed, the red CCD takes images with the i' or z' filters and the CMOS has Y, J, H and Ks filters. For the purposes of the follow-up observations of the Pan-Planets candidates, 3KK was used with the filter selection of g', r', z', H and Ks, as sets of three at a time with either the combination g', z', H or r', z', Ks for a single observation. Since we had already recorded high-precision photometry with the Wendelstein WWFI instrument in the preliminary follow-up, we focused on z' for broader wavelength coverage. 
The used photometric bands were initially chosen to be r', z', Ks but we soon switched the blue optical channel to g' for a more wide-spread wavelength coverage and the NIR channel to H due to signal to noise improvements. The Ks band generally suffers from a large sky background flux and the faintness of our targets means that we deemed the S/N to be insufficient. Furthermore, for true infrared optimization, a telescope would be equipped with an undersized secondary mirror in order to avoid warm light stemming for example from the telescope structure reaching the infrared detector. This is of particular importance for the K band and longer wavelength observations. The FTW is not optimized in this way. To avoid major problems with such long wavelength filters, the FTW utilizes the Ks filter instead of K, which cuts off the red outer edge sooner than the classical K filter.
Nonetheless, there is still a significantly brighter background in the Ks band, which severely affects the signal to noise of fainter sources. We therefore decided to switch to H for our Pan-Planets targets due to their faintness.

We set exposure times to 30\,s and 45\,s, respectively, since longer exposure times would lead to saturation due to the extremely crowded on-sky positions.
Since time-series photometry can be calibrated well with nearby reference stars even at larger sky backgrounds, these candidates were part of Wendelstein's "bad weather" or full moon program. While this led to a vast amount of available data, it simultaneously increased the need for careful image reduction to balance quantity with minimizing noisy data. Between 2017 and 2019, 3KK was used as a survey instrument for these planets. The exact observation dates and image numbers in each filter can be gathered from Tables \ref{tab:pp140dates} and \ref{tab:pp127dates}.
\begin{table}[]
  Wendelstein-1b
  \centering
  \begin{tabular}{c|c|c}
  Date&Bands&Number of frames\\
  \hline
   2015/08/02& (i')& 182 \\  
   2017/08/07& (Ks, r', z')& 452 \\
   2017/08/14 &(Ks, r', z')& 350   \\
   2017/08/22 &(H*,g'*, z'*)& 273  \\
   2018/05/30** &(H, g', z')& 204  \\
   2018/06/03 &(H, g', z')& 266 \\
   2018/06/15** &(H, g', z')& 202 \\
   2018/06/19 &(H*, g', z'*)& 219 \\
   2018/07/29 &(H, g', z')& 247\\
   2018/10/21** &(H, g', z')& 208\\
   &&Total number of images: 2603
  \end{tabular}
  \caption{Observation dates, bands and image numbers for each filter for Wendelstein-1b. *Disregarded due to bad photometric quality. **Secondary eclipse. }
  \label{tab:pp140dates}
\end{table}
\begin{table}[]
Wendelstein-2b
  \centering
  \begin{tabular}{c|c|c}
  Date&Bands&Number of frames\\
  \hline
   2015/07/15& (i') & 106 \\  
   2015/08/06& (i') & 252 \\  
   2017/08/17& (Ks, r', z')& 471 \\
   2017/10/13 &(Ks, r', z')& 170   \\
   2018/05/26 &(H*, g'*, z'*)& 30  \\
   2018/06/02 &(H, g', z')& 220  \\
   2018/06/10** &(H, g', z')& 259 \\
   2018/07/01** &(H, g', z')& 215 \\
   2018/07/08** &(H*, g'*, z'*)& 360 \\
   2018/07/30 &(H, g', z')& 232 \\
   2018/09/25** &(H, g', z')& 200\\
   &&Total number of images: 2515
  \end{tabular}
  \caption{Observation dates, bands and image numbers for each filter for Wendelstein-2b. *Disregarded due to bad photometric quality. **Secondary eclipse. }
  \label{tab:pp127dates}
\end{table}
\subsection{Data reduction and photometry}
The custom Wendelstein 3KK pipeline performs all of the necessary data reduction such as bias and dark subtraction, flat fielding, hot pixel masking, astrometric calibration via the package \textit{scamp} \citep{2006ASPC..351..112B}, crosstalk-correction, nonlinearity-correction, and nightsky-flat subtraction. 
In the following, we give an overview of our process of creating time-series photometry.

Since we observe with 3KK starting at the beginning of its science operations, it was not yet capable of guiding. Instead, a 13-part dither pattern was used with a step size of 8 pixels. That way, we can create so-called night sky flats by masking out stars in every image and then stacking those pattern series into one frame, filling the gaps. This allows for the identification and removal of perturbation sources like hot and cold pixels, stray light and fixed pattern noises by on-sky calibration without the need for any additional overhead. This is especially important for the NIR since the sky background is much higher than in the optical. Dithering and subsequent sky subtraction may introduce higher noise that can degrade image quality. However, since each night has several hundreds of taken images and we use that to self-correct the data, the additional noise is deemed negligible and the small dither pattern keeps other systematics, such as larger-scale flat-field differences, low.

After the frames are satisfyingly calibrated, astrometrically matched and regridded, we performed aperture photometry with a custom python code developed in-house. It is capable of using partial pixels, for example aperture radii that are nonintegers, and we optimized the aperture and parameters and reference stars for each night and filter individually. A starting set of around 15 reference stars was manually chosen, which corresponds to the maximal number of reference stars available with respect to acceptable brightness and distance from the target star. The process was iterated with a varying number of reference stars, sorted from lowest to highest distance from the target star, and variable aperture sizes until the optimal root mean square (RMS) was achieved. 
We determined the RMS by normalizing the time series by creating a smoothed version with a running Gaussian and then measuring the residuals. That way, all of the systematics and photon noise are still kept in the residuals while trends like the transit signal are removed. The iterative selection and deselection of reference stars serves the additional purpose of identifying reference stars that have similar photometric colors compared to the target star. Variable atmospheric extinction during observations introduce color-specific systematics, which were minimized this way and we iterated this process for all used photometric bands to identify band-specific outliers. 
We also included the option of isophotes, where individual masks are created for each frame based on whether the aperture pixels lie above a threshold relative to the sky background, in our case 3$\sigma$ above the sky aperture. This way, an aperture size (and shape) is chosen frame by frame. All these options were iterated and scanned for the best results. 

After the precision was optimized, we corrected our multiband data with an approach similar to \citet{2016AJ....152..171F}. Since we do not want to put in any prior model information, the first step for this was to blindly remove the transit signal. We smoothed the time series photometry with a clipped running Gaussian, dividing the smoothed data by it and then cross-correlated the now flattened multiband data with our implementation of the \textit{sysrem} algorithm \citep{2005MNRAS.356.1466T}, which attempts to identify and eliminate color-dependent systematic effects. The created corrector function was then applied to the presmoothing photometry and its error bars were adjusted. For Wendelstein-1b, this resulted in an average improvement of 20\%, for instance from 5.75\,mmag to 4.78\,mmag in the H-band. The improvement was weaker for Wendelstein-2b with an average improvement of about 14\% due to it being fainter. However, this also serves as a consistency check, since the algorithm is only applicable for systematics. If the data are dominated by photon noise, the algorithm cannot identify the systematic effects as accurately and therefore applies less corrections.
\section{Spectroscopic follow-up}
\subsection{Keck}
We obtained high-resolution optical spectra of Wendelstein-1 using the HIgh Resolution Echelle Spectrometer (HIRES) on the 10m Keck I telescope \citep{1994SPIE.2198..362V} on UT October 01-31 2015. The first observation, recorded only for stellar characterization and verification that no double lines are detectable, was taken without the Iodine cell and confirmed the spectral type. Following that, four more measurements were taken with Iodine. 
We followed the procedures of the California Planet Search \citep[CPS,][]{2010ApJ...721.1467H} and refer to it for further details about the data processing. We used the ''C2'' decker, which provides a spectral resolution of R = 55000, and subtracted the sky from the stellar spectrum. The HIRES spectra were reduced using standard CPS procedures and cover $\sim$3600 -- 8000~\AA. Due to the faint magnitude, the usual approach of creating a template based on the stellar spectra and then cross-correlating each individual image against this template was infeasible. Instead, we used a high S/N template of a similar star previously observed with HIRES and used that as a basis for the RV extraction.

\subsection{HET}
We obtained spectra with the Habitable-zone Planet Finder \citep[HPF,][]{2012SPIE.8446E..1SM}, a temperature stabilized \citep{stefansson2016} fiber-fed NIR spectrograph mounted on the 10\,m Hobby-Eberly Telescope (HET). It operates in the wavelength region of 810 - 1280 nm at a resolution of 55000. Due to HET's large 10\,m-class mirror with an effective aperture size of 9.2\,m and HPF's excellent performance for objects with their emission maximum in the infrared, it is the ideal combination for this follow-up operation. Exposures at HPF have to be split into blocks no longer than 945\,s, therefore we split observations into blocks of two or three. We recorded reconnaissance spectroscopy for Wendelstein-2b in June 2018 and, after verifying that an eclipsing binary is highly unlikely both based on multiband photometry and the initial RV data, performed a more comprehensive observing campaign in August-October 2019. Excluding overheads, we recorded six additional data points per target with up to 30\,min of exposure time per point, depending on the weather conditions and constraints based on HET's design based on a fixed-angle primary mirror. Scheduling was performed automatically by the HET observers with the goal of spreading the observations evenly across the planets' phases. 

The HPF spectra were reduced and RVs extracted using the procedures described in \cite{stefansson2019}. In short, the HPF 1D spectra were reduced and extracted with the custom HPF data-extraction pipeline described further in \cite{ninan2018}, \cite{kaplan2018}, and \cite{2019Optic...6..233M}. Following the spectral extraction, the RVs were extracted using a modified version of the SpEctrum Radial Velocity Analyzer (SERVAL) pipeline \citep{zechmeister2018}, optimized for the HPF spectra. Our modifications to SERVAL to adapt it to HPF spectra are further described in \cite{2019Optic...6..233M} and \cite{stefansson2019} . This pipeline uses the \texttt{barycorrpy} Python package \citep{kanodia2018} to calculate barycentric corrections, which relies on the barycentric-correction algorithms from \cite{wright2014}. The HPF spectra are wavelength-calibrated and corrected for RV drift based on calibration observations taken with the dedicated HPF Laser Frequency Comb (LFC) during dedicated evening, morning, and periodic calibration exposures taken throughout the night to monitor the HPF drift. The HPF LFC and its performance is described in detail in \cite{2019Optic...6..233M} and \cite{stefansson2019} .

\section{Verification}
\subsection{Radial velocity measurements}
\label{section:RV}
\begin{figure}
  \centering
  \includegraphics[width=\linewidth]{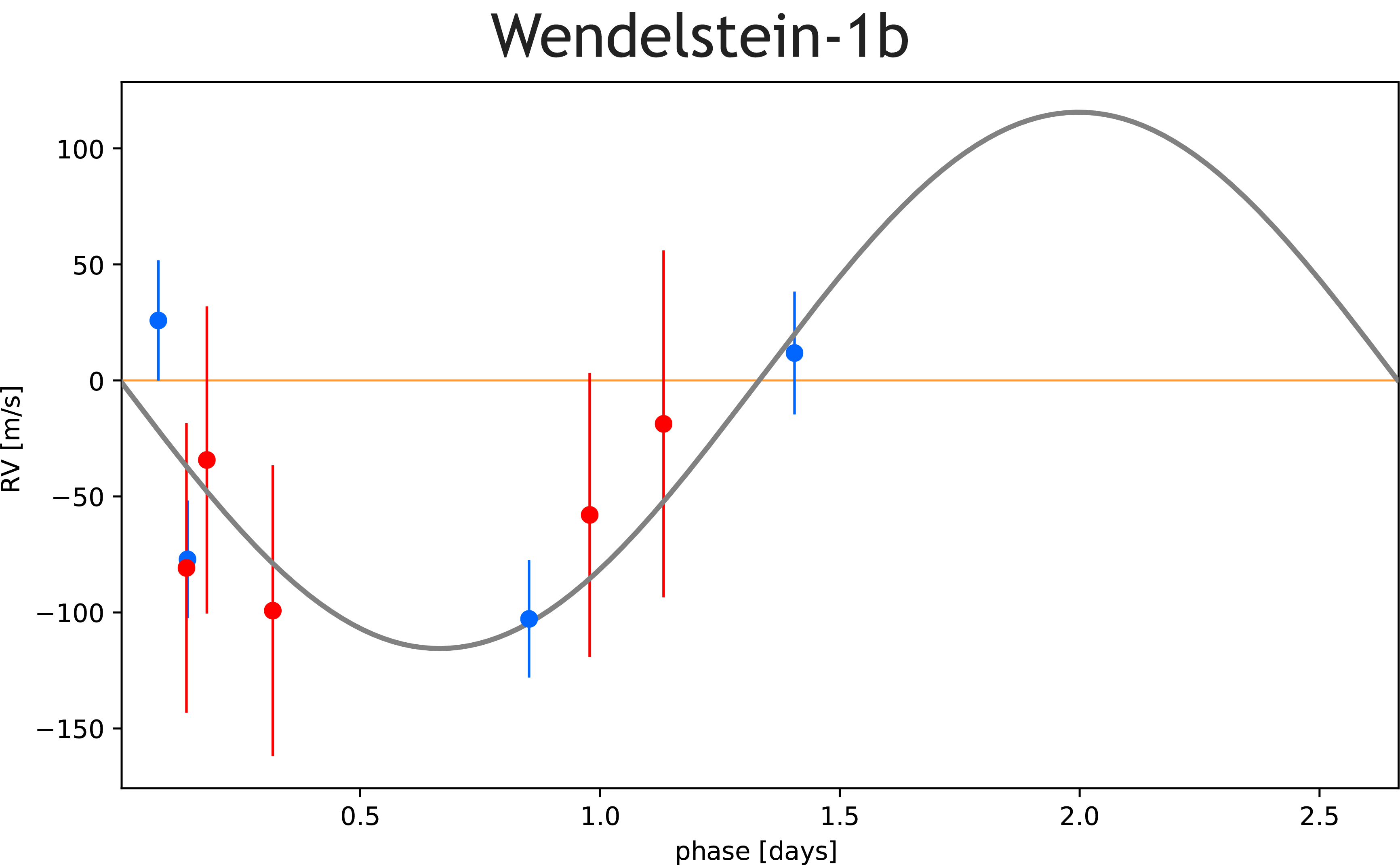}
  \caption{Orbital solution of Wendelstein-1b, showing the best-fitting RV function phase-folded to its period and all measured RV points.}
  \label{fig:PP140_14711b_rv}
\end{figure}

\begin{figure}
  \centering
  \includegraphics[width=\linewidth]{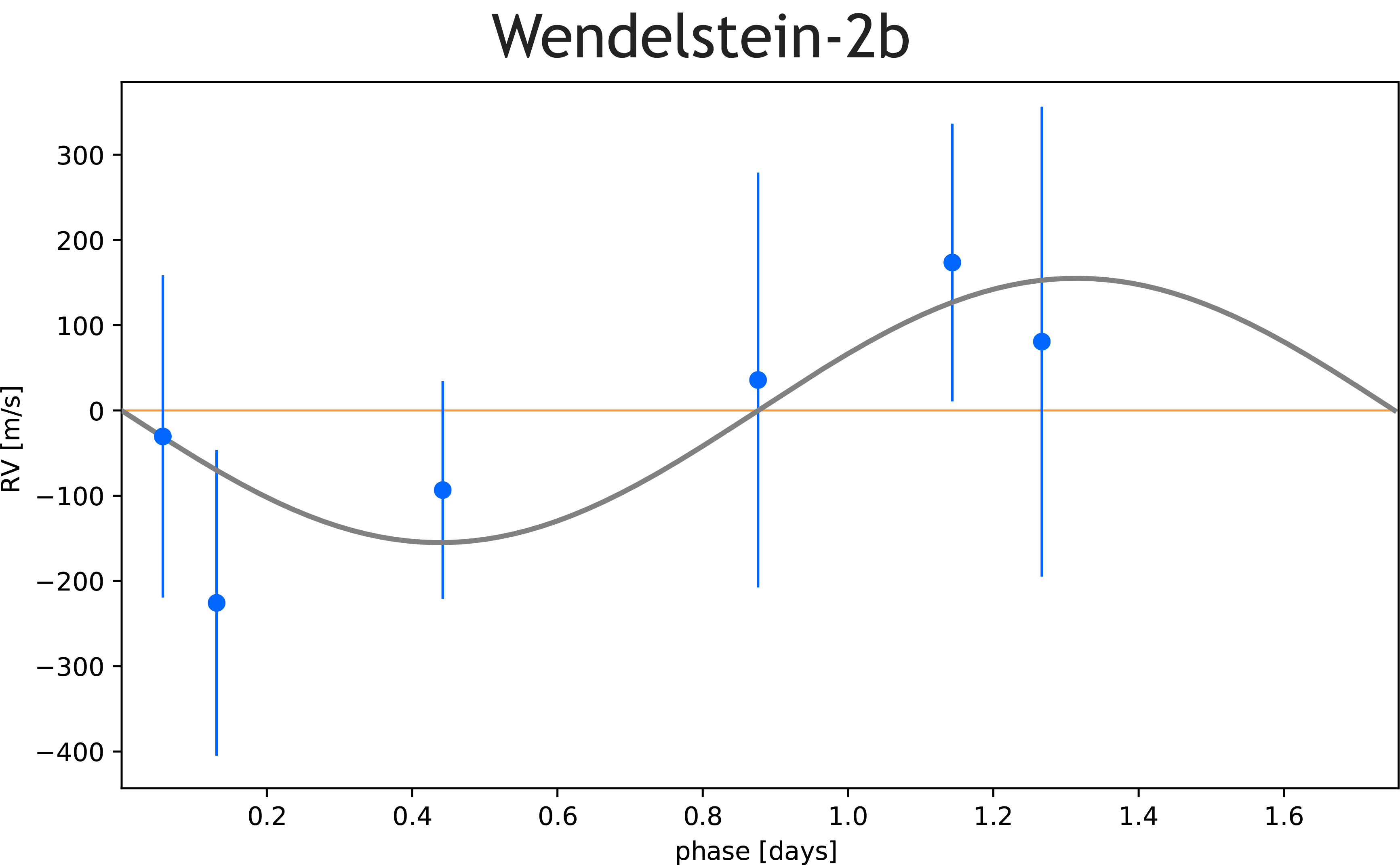}
  \caption{Orbital solution of Wendelstein-2b, showing the best-fitting RV function phase-folded to its period and all measured RV points.}
  \label{fig:PP127_21645b_rv}
\end{figure}

We fit the observed radial velocity data with the \textit{Exostriker} code \citep{2019ascl.soft06004T} and independently of the multiband photometry, since no code is yet capable of doing so. For Wendelstein-1, we calibrated the Keck and HPF data for their mutual offset by keeping the systemic velocity as a free parameter and performed a joint fit, while Wendelstein-2 only has data from HPF. We estimated the stellar jitter for both systems and used that to adjust the measured RV uncertainties based on the data reduction procedures. We disregarded three data points from the HPF measurements, since those were recorded and flagged in bad nights where the signal to noise was unfavorable and observations were stopped before the sought three 900\,s exposures. The fitting results are shown in Fig. \ref{fig:PP140_14711b_rv} for Wendelstein-1b and Fig. \ref{fig:PP127_21645b_rv} for Wendelstein-2b, while the resulting radial velocity data are displayed in Tables \ref{tab:pp140-14711-rv} and \ref{tab:pp127-21645-rv}, respectively. 
\begin{table}[hbt]
  \centering
  \begin{tabular}{c|c|c|c} 
  & Wendelstein-1 & \\
  \hline
    BJD (TDB) & RV [m/s] & e$_{RV}$ [m/s] & Instrument \\
     \hline
 2457296.89886 & -100 & 22&HIRES\\
 2457298.78952 &29 & 22 &HIRES\\
 2457298.85034 & -74 & 22 &HIRES\\
 2457326.74985 & 15 & 23 &HIRES\\
 2458702.65045 & -99 & 63 &HPF\\
 2458707.83983 & -34 & 66 &HPF\\
 2458708.63780 & -58 & 61 &HPF\\
 2458739.75838 & -81 & 62 &HPF\\
 2458740.75289 & -19 & 75 &HPF\\
  \end{tabular}
  \caption{Radial velocity measurements taken for Wendelstein-1 and their respective errors, corrected for stellar jitter, and used instrument.}
  \label{tab:pp140-14711-rv}
\end{table}
 \begin{table}[hbt]
  \centering 
  \begin{tabular}{c|c|c|c} 
  & Wendelstein-2 & \\
  \hline
    BJD (TDB) & RV [m/s] & e$_{RV}$ [m/s] & Instrument \\
     \hline
2458289.77660 & 35 & 243 &HPF\\
2458290.78361 & -226 & 179 &HPF\\
2458320.88305 & -94 & 128 &HPF\\
2458428.59289 & 80 & 276 &HPF\\
2458708.82517 & 173 & 163 &HPF\\
2458728.76542 & -31 & 189 &HPF\\
  \end{tabular}
  \caption{Radial velocity measurements taken for Wendelstein-2 and their respective errors, corrected for stellar jitter, and used instrument.}
  \label{tab:pp127-21645-rv}
\end{table}

In systems with visible transits, the fitting process is significantly easier due to very precise estimates for period and t$_0$. Since we have almost 10 years of photometric data, we therefore held those values fixed and only kept the amplitude K as a free parameter. Due to the very short periods of the systems, eccentricity $e$ is likely to be close to zero and the photometric data had no indication of a largely eccentric orbit. The low number of RV points further hinderd eccentricity measurements, but instead of defaulting it to $e=0$, we performed two distinct fits, one with $e$ as a free parameter and one with it being fixed to 0. For Wendelstein-1b, the free fit indicates a circular orbit with the fit value for $e=0.0198^{+0.113}_{-0.0198}$ while for Wendelstein-2b, the best fit favours a higher eccentricity of $e=0.193^{+0.446}_{-0.193}$. However, the low number of RV points makes any conclusion problematic. We therefore assumed that both planets are on circular or almost-circular orbits where we can only provide upper limits to their eccentricities.

Although the results from RV are limited by the low number of data, the precisely measured transit periods mean that any scenario involving a brown dwarf or stellar companion can be ruled out completely. While a null result is statistically possible with p=0.012 and p=0.201 for Wendelstein-1b and Wendelstein-2b, respectively, this only means that the RV signal cannot be distinguished well enough from a zero-mass case. The opposite case of having a stellar companion would result in RV amplitudes in the order of 2000-10000 m/s and this can be ruled out conclusively. However, the measured mass of Wendelstein-2b in particular has a large uncertainty, so it is not clear whether those systems are hot Jupiters or Saturns. Further observations will have to be taken to improve those estimates. The final systems parameters can be found in Tables \ref{tab:pp140_fitpara} and \ref{tab:pp127_fitpara}, respectively.
\subsection{Transit color signature}
\label{sec:transit_color_sig}
Multiband photometry creates a pathway toward improving planetary parameter fits by providing new degrees of freedom for limb darkening (LD) coefficients while simultaneously allowing one to constrain other parameters, such as inclination and the radius ratio. Besides attaining better parameters, the idea to use Transit Color Signature (TraCS) as a way of confirming a transit signal is by no means new. \citet{Rosenblatt71} theorized about this approach in the very first publication suggesting the transit method to detect exoplanets. Based on that, a transit event of a planet and that of an eclipsing binary may look indistinguishable when observed in just one photometric band, but begin to exhibit unique features setting them apart from one another once observed in multiple bands. This offers an alternative pathway to planet confirmation that can be more efficient in terms of time and cost. This is particularly applicable for very faint stars, as are the two candidates discussed in this work, since they are extremely challenging to follow-up with high-resolution spectroscopy even when using the largest telescopes to date as shown in Sect. \ref{section:RV}.

Multiband photometry uses two main circumstances to its advantage. Considering the case of a grazing eclipse by a stellar companion, limb darkening of the stellar host causes the transit depth of an eclipsing binary transit to increase from shorter to longer wavelengths used in observations. The occultation caused by a planetary transit however, will in most cases not change in depth much throughout all wavelengths if the orbit inclination is closer to 90$^{\circ}$, the reason for this being the increase of severity of the limb darkening effect for short wavelengths. 
Longer wavelengths, in contrast, are left fairly unchanged by change in temperature and only show marginal change in spectral radiance for cooler or hotter origins of photons. Hence, transiting companion stars will block less flux in a relative sense at shorter wavelengths than they do at longer wavelengths. A planet is always moving across the entire stellar surface due to inclination constraints for it to be observed at all. 

There are other scenarios where the above case is not true, for example a distant eclipsing binary system that is blended in the PSF of a foreground star. Here, multiband can provide a way to disentangle the scenarios if the stellar types of foreground and background systems differ, which results in different eclipse depths in relation to wavelength.

However, this is still not enough to safely confirm a system, unless one measures the companion's effective temperature during secondary eclipse, as we suggest to do in the TraCS technique and have done for the planets in this work. This way, even highly unlikely but possible scenarios such as a blended EB system where the primary star has the same stellar type as the foreground star can be ruled out conclusively.

This method provides the proficiency to discern a true signal from a false positive in a cheap and time-efficient manner, allows for rapid data analysis and, with sufficiently good quality observations, a good confidence level for false positive detection. Its true strengths lie in the small mirror sizes that are necessary for photometry and the possibility of upgrading old 1\,m-class telescopes with NIR capabilities.

\subsection{Multiband fit}
In our simultaneous multiband fit, the parameters for period $p$, initial transit time $t_0$, inclination $i$, semi-major axis $a$, eccentricity $e$ and radius ratio $R_p$ were kept fixed across all photometric bands, while LD coefficients were fit individually and assigned priors according to stellar parameter uncertainties. The fitting wrapper is being developed in-house and utilizes the \textit{emcee} code by \cite{2013PASP..125..306F} for MCMC, \textit{batman} by \cite{2015PASP..127.1161K} for transit-model fitting and \textit{ldtk} \citep{2015MNRAS.453.3821P} for estimating physical LD priors. We further modified \textit{ldtk} by improving error propagation and forwarding the internal LD prior function into our fit-package as function object that can be evaluated for any LD parameter.

Wendelstein-1b and Wendelstein-2b are both true planet detections with robust and consistent fitting results across all wavelengths. Wendelstein-2b shows a notably deeper transit in the r' band at a confidence of 4$\sigma$, which may be indicative of an $H_\alpha$ envelope that, due to spectral broadening, affects broad-band photometry as well. \citet{2018NatAs...2..714Y} measured a hydrogen envelope of 1.64 planetary radii for KELT-9b and while Wendelstein-2b is significantly cooler at about $1852_{-140}^{+120}$\,K, it has a very close orbit of $0.0234\pm 0.0015$\,a.u. and may also have an extended atmosphere. Further observations with narrow-band filters are planned with 3KK to observe this anomaly in greater detail.

Comparing the determined quadratic limb darkening coefficients to \citet{2011A&A...529A..75C} and \citet{2012yCat..35460014C}, we can conclude that Wendelstein-1 is indeed about 200\,K cooler than Wendelstein-2 and that the fits are consistent with stars of log(g) = 4.5 and effective temperatures of 4200\,K $\pm$ 200\,K and 4400\,K $\pm$ 200\,K, respectively. As a final verification step, we reran the fit without any priors on limb darkening and the results were consistent with our prior-based ones. We have therefore been able to independently verify the stellar types of our candidates. 

\begin{table}[]
  \centering
  Wendelstein-1b\\
  \begin{tabular}{c|c}
   \hline
    \textbf{Parameter} & \textbf{Result}\\ \hline
     \textit{P} [d] & $2.663416 \pm 0.000001$ \\
     \textit{t0} [BJD$\_$TBD] & $2455367.738464 \pm 0.000014$ \\
     \textit{$R_P$} [R$_{*}$]& $0.1698 \pm 0.0010$ \\ 
     \textit{a} [a.u.] & $0.0282 \pm 0.0015$\\ 
     \textit{i} [deg] & $86.12_{-0.39}^{+0.43}$ \\ 
     \textit{$e$}* & $ < 0.056$ \\
     \textit{$e$}**& $< 0.012$ \\
     \textit{M} [M$_J$] & $0.59_{-0.13}^{+0.17}$ \\
     \textit{$\rho$} [g/cm$^3 ] $    &   $0.72_{-0.17} ^{+0.23}$      \\
     \textit{K} [m/s] & $115.56 \pm 32.59$ \\
     \textit{T$_{pl}$} [K] & $\leq 1884 (1\sigma); 2198 (3\sigma)$\\
     \textit{$ \Delta $ F(g')} & 0.0223 $ \pm $ 0.0009\\
     \textit{$ \Delta $ F(r')} & 0.0257 $ \pm $ 0.0006\\
     \textit{$ \Delta $ F(i')} & 0.0269 $ \pm $ 0.0003\\
     \textit{$ \Delta $ F(z')} & 0.0269 $ \pm $ 0.0010\\
     \textit{$ \Delta $ F(H)} & 0.0275 $ \pm $ 0.0007\\
     \textit{$ \Delta $ F(Ks)} & 0.0281 $ \pm $ 0.0011\\
     \textit{$u_1$(g')} & $0.89_{-0.02}^{+0.02}$\\ 
     \textit{$u_2$(g')} &$-0.07_{-0.02}^{+0.02}$ \\ 
     \textit{$u_1$(r')} & $0.68_{-0.01}^{+0.01}$ \\ 
     \textit{$u_2$(r')} & $0.07_{-0.01}^{+0.01}$ \\ 
     \textit{$u_1$(i')} &$0.50_{-0.01}^{+0.01}$ \\ 
     \textit{$u_2$(i')} &$0.15_{-0.01}^{+0.01}$ \\
     \textit{$u_1$(z')} & $0.41_{-0.02}^{+0.02}$ \\ 
     \textit{$u_2$(z')} & $0.17_{-0.02}^{+0.02}$ \\ 
     \textit{$u_1$(H)} & $0.23_{-0.02}^{+0.02}$ \\ 
     \textit{$u_2$(H)} & $0.22_{-0.02}^{+0.02}$ \\ 
     \textit{$u_1$(Ks)} & $0.18_{-0.02}^{+0.02}$ \\ 
     \textit{$u_2$(Ks)} & $0.20_{-0.02}^{+0.02}$ \\ 
    \hline
  \end{tabular}
  \caption{Orbit parameters and posteriors of Wendelstein-1b, based on separate multiband photometry and RV fitting. $ \Delta $ F denotes the flux drop for the corresponding band. The uncertainties were determined via the MCMC processes as described in Sects. \ref{section:RV} and \ref{sec:transit_color_sig}. The quadratic limb darkening coefficients $u_1$ and $u_2$ are given for each photometric band. *upper limit based on transit measurements; **upper limit based on RV measurements}
  \label{tab:pp140_fitpara}
\end{table}
\begin{table}[]
  \centering
 Wendelstein-2b\\
  \begin{tabular}{c|c}
   \hline
    \textbf{Parameter} & \textbf{Result}\\ \hline
     \textit{P} [d] & $ 1.7522239 \pm 0.0000008$ \\ 
     \textit{t0} [BJD$\_$TBD] & $2455679.254400 \pm 0.000013 $ \\
     \textit{$R_P$} [R$_{*}$]& $0.176 \pm 0.004$ \\ 
     \textit{a} [a.u.] &  $0.0234 \pm 0.0015$\\ 
     \textit{i} [deg] & $87.87_{-1.04}^{+0.98}$ \\ 
    \textit{$e$}* & $< 0.057$ \\
     \textit{$e$}**& $< 0.193$ \\
     \textit{M} [M$_J$] & $0.73_{-0.32}^{+0.55}$ \\
     \textit{$\rho$} [g/cm$^3 ] $    &   $0.62_{-0.26}^{+0.46} $      \\
     \textit{K} [m/s] & $155.584 \pm 66.116$ \\
     \textit{T$_{pl}$} [K] & $\leq 2158 (1\sigma); 2470 (3\sigma)$ \\
     \textit{$ \Delta $ F(g')} & 0.0355 $ \pm $ 0.0014 \\
     \textit{$ \Delta $ F(r')} & 0.0397 $ \pm $ 0.0011 \\
     \textit{$ \Delta $ F(i')} & 0.0359 $ \pm $ 0.0004 \\
     \textit{$ \Delta $ F(z')} & 0.0355 $ \pm $ 0.0009 \\
     \textit{$ \Delta $ F(H)} & 0.0366 $ \pm $ 0.0011 \\
     \textit{$ \Delta $ F(Ks)} & 0.0375 $ \pm $ 0.0022 \\
     \textit{$u_1$(g')} & $0.96_{-0.02}^{+0.02}$\\ 
     \textit{$u_2$(g')} &$-0.13_{-0.02}^{+0.02}$ \\ 
     \textit{$u_1$(r')} & $0.70_{-0.01}^{+0.01}$ \\ 
     \textit{$u_2$(r')} & $0.04_{-0.01}^{+0.01}$ \\ 
     \textit{$u_1$(i')} &$0.54_{-0.01}^{+0.01}$ \\ 
     \textit{$u_2$(i')} &$0.10_{-0.01}^{+0.01}$ \\
     \textit{$u_1$(z')} & $0.46_{-0.02}^{+0.02}$ \\ 
     \textit{$u_2$(z')} & $0.12_{-0.02}^{+0.02}$ \\ 
     \textit{$u_1$(H)} & $0.21_{-0.02}^{+0.02}$ \\ 
     \textit{$u_2$(H)} & $0.24_{-0.02}^{+0.02}$ \\ 
     \textit{$u_1$(Ks)} & $0.18_{-0.02}^{+0.02}$ \\ 
     \textit{$u_2$(Ks)} & $0.18_{-0.02}^{+0.02}$ \\ 
    \hline
  \end{tabular}
  \caption{Orbit parameters and posteriors of Wendelstein-2b, based on separate multiband photometry and RV fitting. $ \Delta $ F denotes the flux drop for the corresponding band. The uncertainties were determined via the MCMC processes as described in sections \ref{section:RV} and \ref{sec:transit_color_sig}. The quadratic limb darkening coefficients $u_1$ and $u_2$ are given for each photometric band. *upper limit based on transit measurements; **upper limit based on RV measurements}
  \label{tab:pp127_fitpara}
\end{table}

\begin{figure*}
  \centering
  \includegraphics[width=\linewidth]{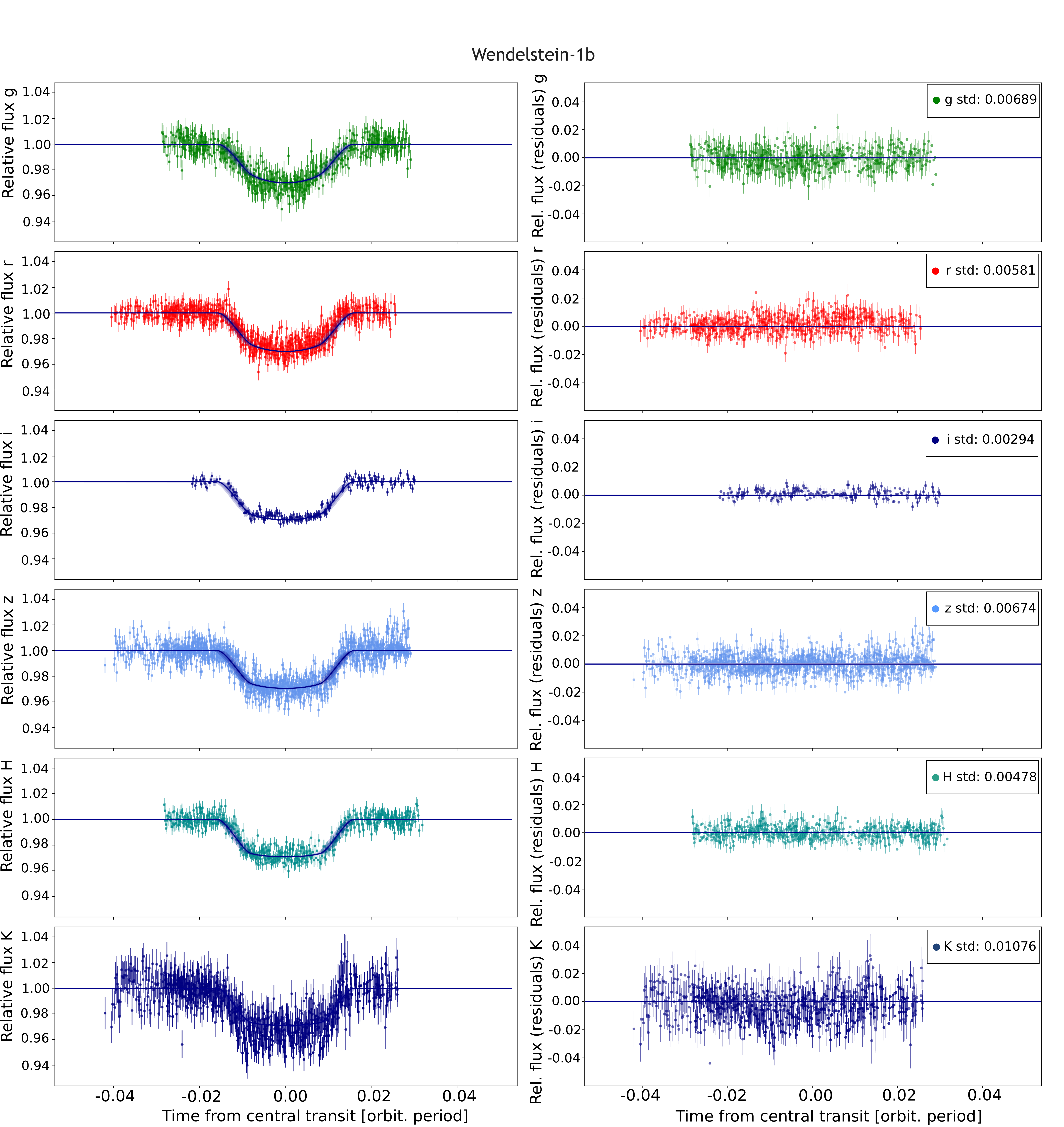}
  \caption{Multiband fit of Wendelstein-1b in the photometric bands g', r', i', z', H, Ks sorted by mean wavelength with the data collected by WWFI and 3KK and their corresponding best fit. Uncertainties of the fit are shown as transparent shades for 1$\sigma$, 2$\sigma$ and 3$\sigma$. The residuals for the fit are shown in the right panel. }
  \label{fig:PP140_MCMC}
\end{figure*}
\begin{figure*}
  \centering
  \includegraphics[width=\linewidth]{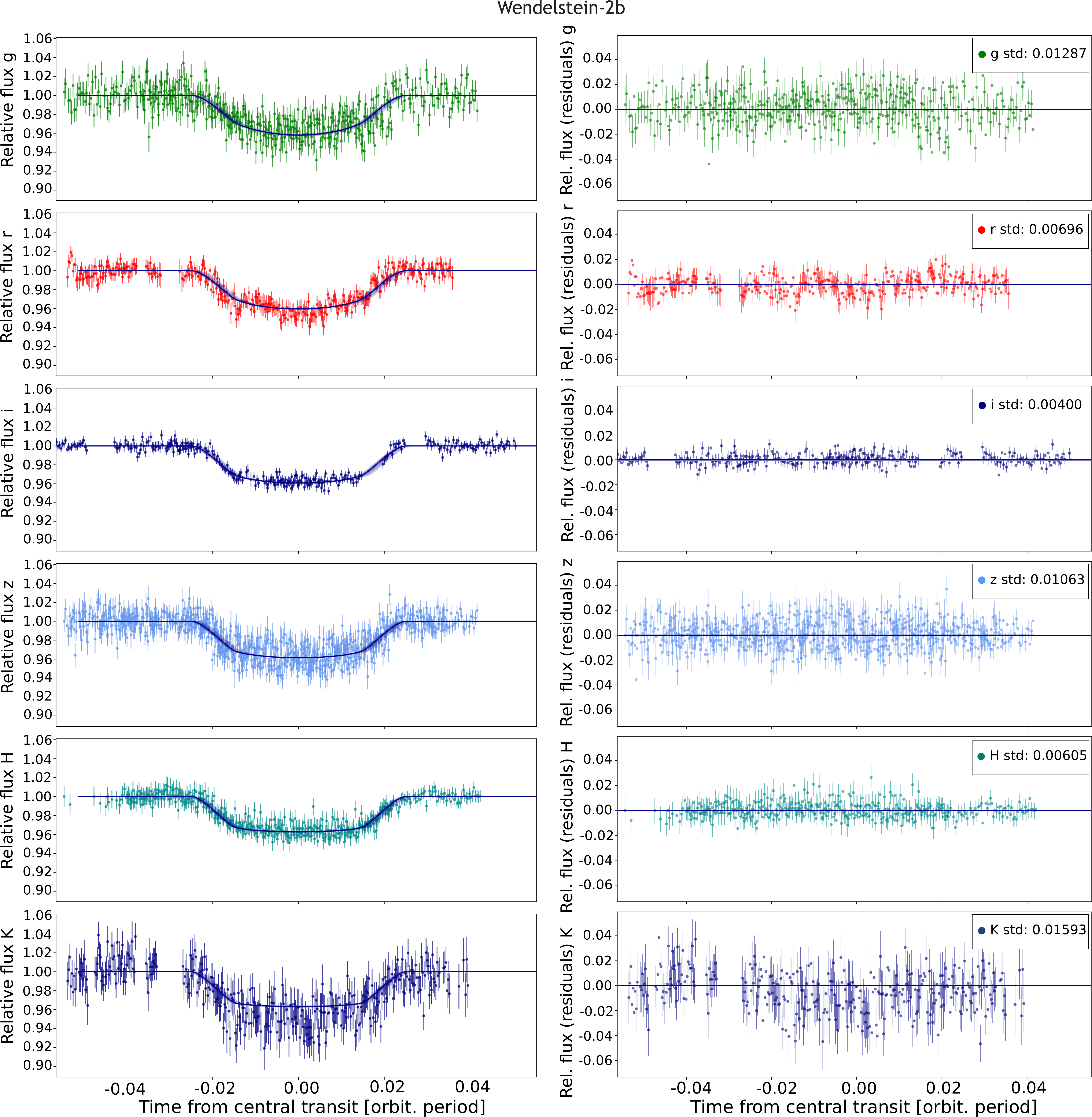}
  \caption{Multiband fit of Wendelstein-2b in the photometric bands g', r', i', z', H, Ks sorted by mean wavelength with the data collected by WWFI and 3KK and their corresponding best fit. Uncertainties of the fit are shown as transparent shades for 1$\sigma$, 2$\sigma$ and 3$\sigma$. The residuals for the fit are shown in the right panel. }
  \label{fig:PP127_MCMC}
\end{figure*}

\subsection{Potential contamination}
\label{sec:pot_contamination}
Observed stars, particularly when located close to the galactic disk, may be a blend of two or more stellar sources. Such a blend will dilute the overall transit depth, leading to wrong system parameter estimates or even misidentification of eclipsing binaries as transiting planets. While this effect is already in part accounted for by multiband transit photometry, a nearby blend with similar color could show no varying light dilution across the photometric bands. 

\begin{figure}
  \centering
  \includegraphics[width=0.9\linewidth]{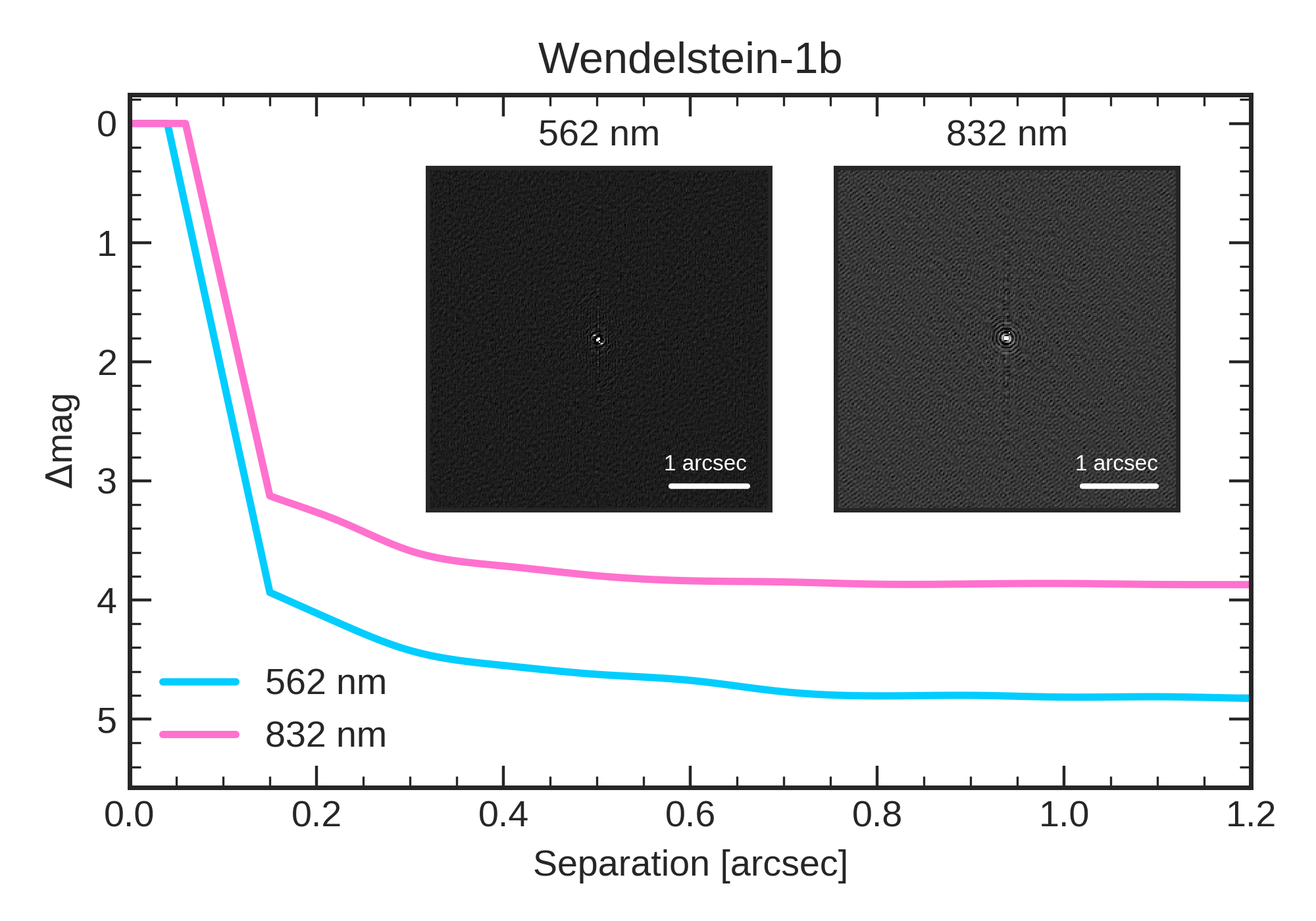}
  \caption{Speckle imaging of Wendelstein-1 with the reconstructed images and contrast curve for both observed bands. The star appears single in both images and the sensitivity curves rule out brighter close companions or background stars that would contribute significant flux to the transit light curve.}
  \label{fig:Wendelstein-1-speckle}
\end{figure}

Following \citet{2018AJ....156...78L}, we conducted high-resolution speckle imaging of Wendelstein-1b on UT 2017 May 13 to further rule out the possibility of contaminating sources, using the NASA Exoplanet Star and Speckle Imager \citep[NESSI;][]{2018PASP..130e4502S} mounted on the WIYN 3.5m telescope at Kitt Peak. NESSI uses high-speed electron-multiplying CCDs to provide simultaneous imaging at 25\,Hz in two 44\,nm-wide bands centered at 562\,nm and 832\,nm. The data are acquired and reduced following \citet{2011AJ....142...19H}, yielding 4.6''$\times$4.6'' reconstructed images. Fig. \ref{fig:Wendelstein-1-speckle} shows the resulting images and 5$\sigma$ contrast curves, which exclude contaminants down to 4\,mag (3\,mag) at 562\,nm (832\,nm) at an angular separation of 0.2\,arcsec. 

The problem with Wendelstein-2b as an example of faint planet candidates is that our attempts to attain high-angular precision imaging failed due to insufficient signal to noise. We therefore used another pathway, the centroiding technique, which is by now well-established and used by different projects \citep{2010ApJ...713L.103B,2017MNRAS.472..295G,2018ApJ...869L...7A}. By measuring the center of gravity of the analyzed star, any offset during the transit may indicate a visual companion that is hidden in the PSF. Having such a companion dilutes the depth of the transit, which may lead to a false-positive identification of an eclipsing binary as a planet, a scenario that has to be ruled out.

We implemented the center of gravity measurement during aperture photometry and calibrated the target star's center with the previously selected photometric reference stars. In addition to that, we constructed a simulation based on the image's PSF, detector pixel scale, sky background noise, photon noise and used apertures. Then, we simulated the primary star and injected a second star for which we varied its brightness and distance from the primary. Those images were then fit with the same centroiding module from the aperture photometry code and we determined both relative shifts of the center of gravity and the uncertainties when a transit with the measured brightness drop occurs. We further correlated the uncertainties with the frame-by-frame centroid variations. 
\begin{figure}
  \centering
  \includegraphics[width=0.9\linewidth]{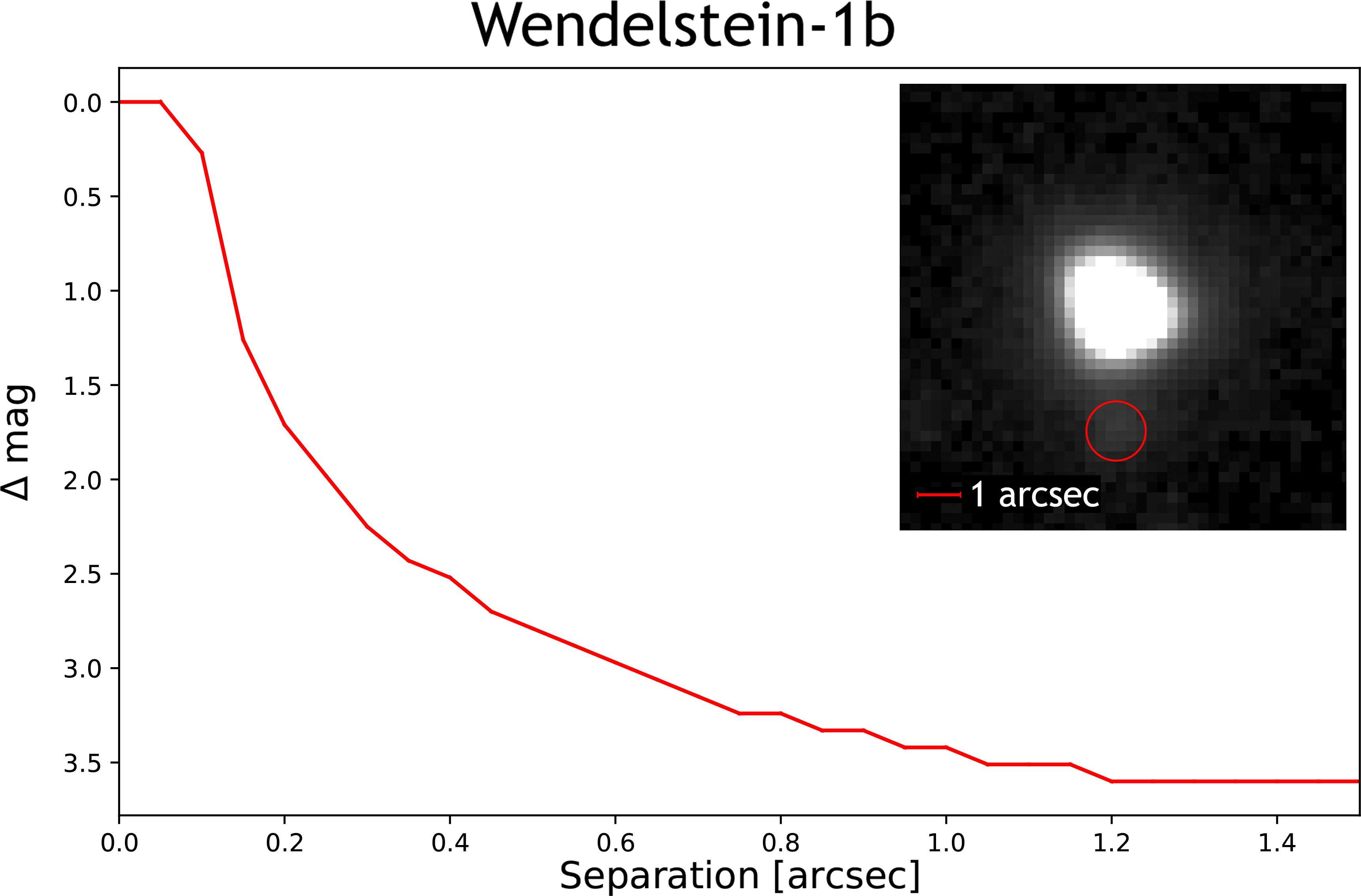}
  \caption{Contrast curve based on differential centroid measurements before and during Wendelstein-1b's transit for the H band. An image cutout of the star is shown in the top right, extracted from a stacked reference frame that has an overall exposure time of 6367\,s. We identify a potential contaminant, marked by a red circle, measured about 6\,mag fainter in the H-band at a separation of about 3\,arcsec, which was not detected in the Gaia survey. The slight deformation of the star is due to a telescope misalignment which affected all stars in the FOV. }
  \label{fig:Wendelstein-1-centroid}
\end{figure}

\begin{figure}
  \centering
  \includegraphics[width=0.9\linewidth]{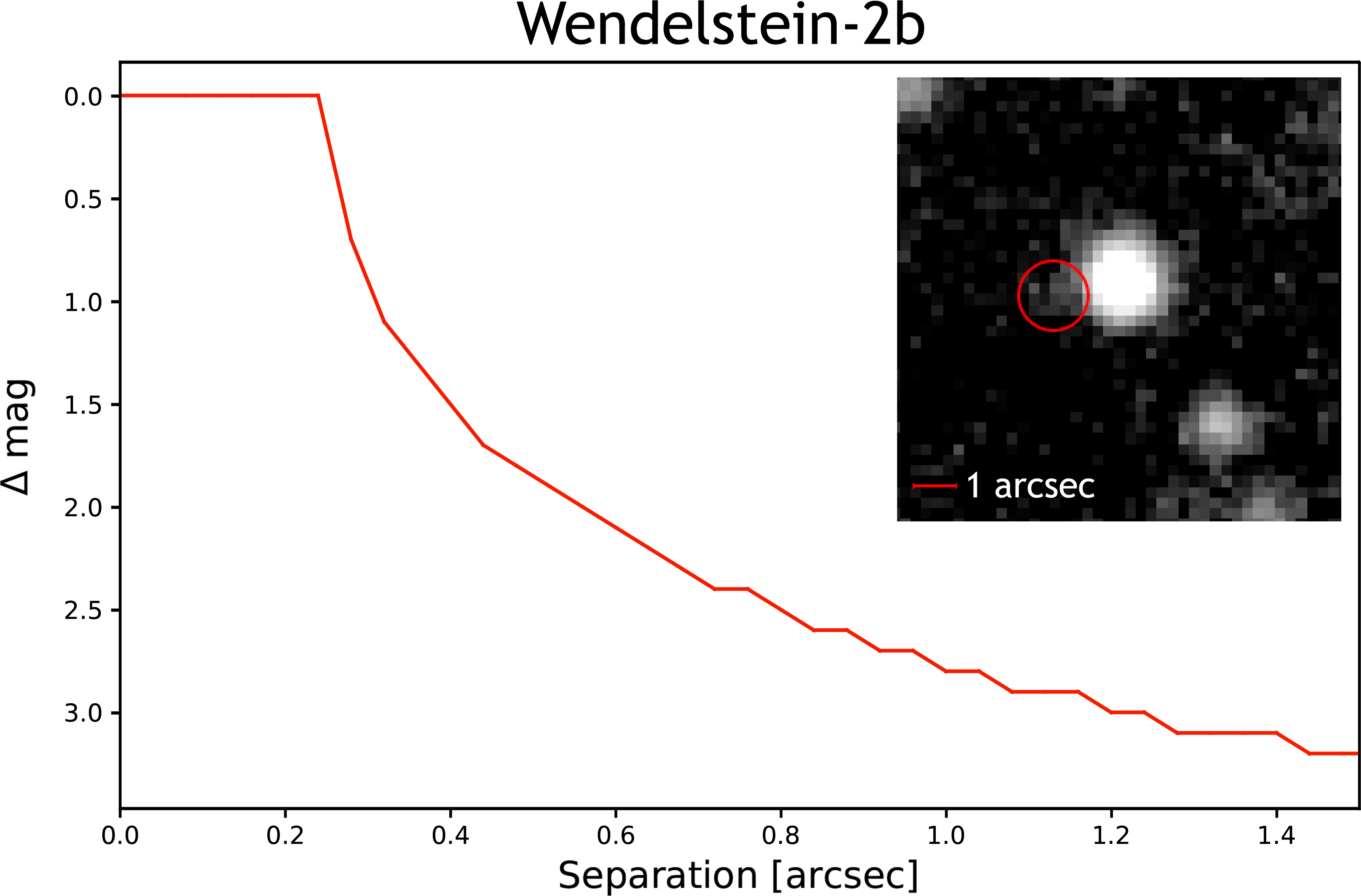}
  \caption{Contrast curve based on differential centroid measurements before and during Wendelstein-2b's transit for the H band. An image cutout of the star is shown in the top right, extracted from a stacked reference frame that has an overall exposure time of 3438\,s. The potential contaminant, identified in the Gaia survey, is marked by a red circle based on the given RA/DEC coordinates.}
  \label{fig:Wendelstein-2-centroid}
\end{figure}

The centroids were measured with an average accuracy of 0.04" and 0.06" for Wendelstein-1b and Wendelstein-2b, respectively. For Wendelstein-2b, a small in-transit shift of about 0.15" was detected at wider apertures, a first indication that a fainter star may be present. Based on our simulations, we created contrast curves shown in Figs. \ref{fig:Wendelstein-1-centroid} and \ref{fig:Wendelstein-2-centroid} similar to classical high angular resolution imaging. Measurements for Wendelstein-2b resulted in an upper limit of 3.2\,mag at a separation of 1.4\,arcsec. Coincidentally, a faint source\footnote{Gaia ID 1820764459338195840} has been detected by Gaia at a separation of about 1.35\,arcsec and about 3.9\,mag fainter so it is highly likely that there is a slight contamination of Wendelstein-2b. We could also identify this source in our H-band reference frame after stacking 3438\,s of our sharpest images, as is shown in Fig. \ref{fig:Wendelstein-2-centroid}. As a consequence, we reran aperture photometry while masking out the contaminant's region with no detectable impact on the transit depth, so we assumed that no dilution is taking place. For Wendelstein-1b, it confirms the contrast curve attained from speckle imaging, albeit at a marginally lower magnitude threshold. In the stacked H-band reference frame, we identified a faint source about 6\,mag fainter at a separation of about 3\,arcsec. This source was not detected in the Gaia survey and its large angular separation means that contamination is not possible.

\subsection{Secondary eclipses}
\begin{figure}
  \centering
  \includegraphics[width=\linewidth]{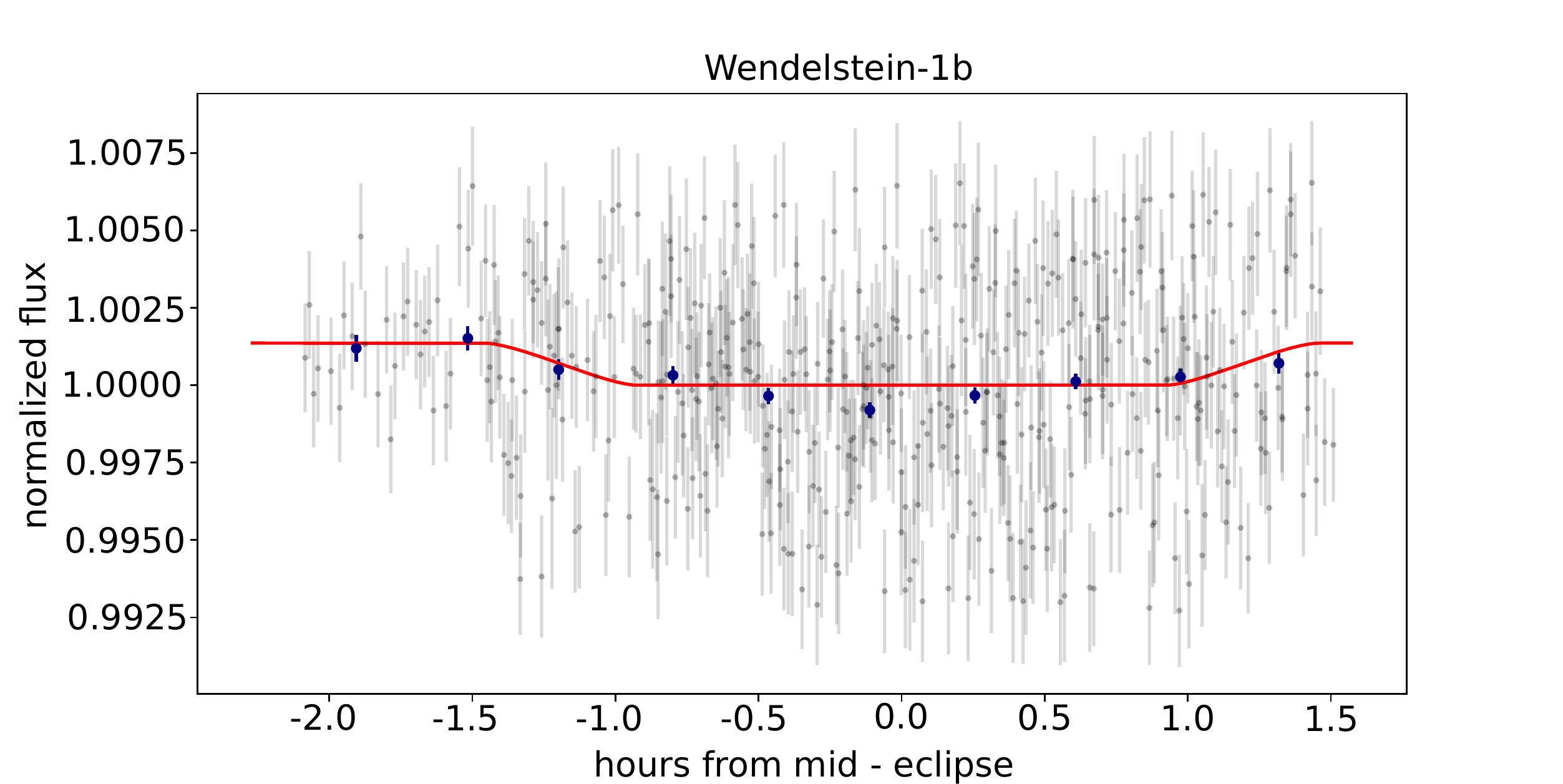}
  \caption{Folded light curve of Wendelstein-1b's secondary eclipse measurements with an out of eclipse shift of 0.00136$\pm$ 0.00034. The original data points with their error bars are shown in gray, the best-fitting secondary eclipse model as a red line. For better visibility, the data points are also shown as blue bins with 15\,min intervals.}
  \label{fig:PP140_sec}
\end{figure}

\begin{figure}
  \centering
  \includegraphics[width=\linewidth]{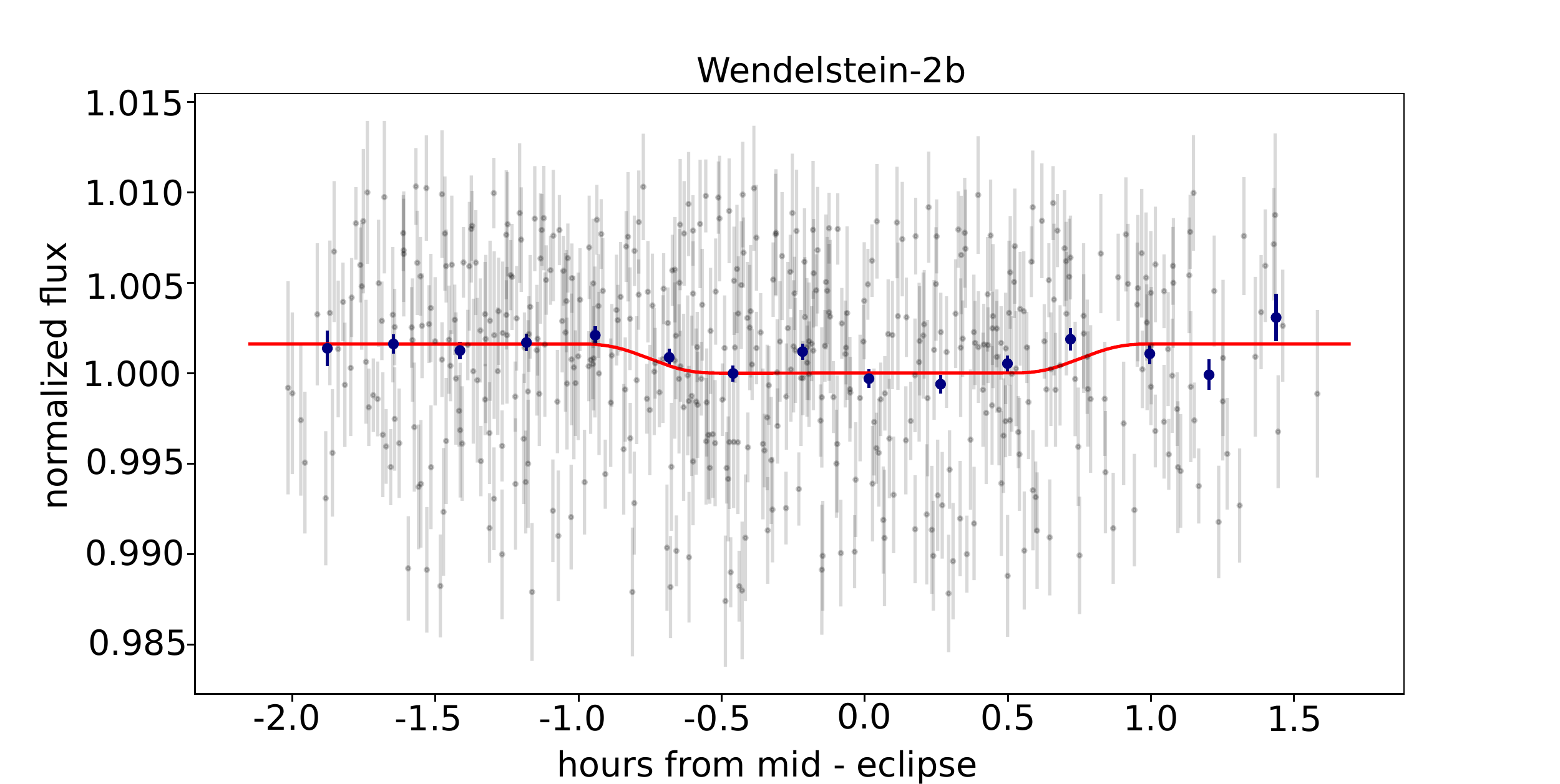}
  \caption{Folded light curve of Wendelstein-2b's secondary eclipse measurements with an out of eclipse shift of 0.00161$\pm$ 0.00041. The original data points with their error bars are shown in gray, the best-fitting secondary eclipse model as a red line. For better visibility, the data points are also shown as blue bins with 15\,min intervals.}
  \label{fig:PP127_sec}
\end{figure}

Ground-based observations of secondary eclipses are a challenging task, even more so with faint targets like Wendelstein-1b and Wendelstein-2b. Fainter stars lie on the same order of magnitude in flux as the sky background in Ks. To optimize for the detection of these events, we therefore decided to use the H band instead of the more intuitive Ks band due to the vastly better photometric accuracy found in the primary transits. 
The recorded measurements of the predicted secondary eclipse phase-point serve two purposes. The primary goal is the exclusion of any remaining false-positive detection scenarios, for example by a Jupiter-sized, low-mass star that would be detectable by a significant secondary eclipse depth. We improved the quality of the light curve during the secondary eclipse with repeated H-band photometry to reach the secondary goal of determining the planet's temperature. 
For this, we again used the \textit{batman} code by \cite{2015PASP..127.1161K} where we kept two free parameters for the fit: firstly, the planet/star flux ratio and secondly the phase offset of the secondary eclipse that is affected by eccentricity and the argument of periastron.
Since this is numerically easier due to only two free parameters, we simply constructed a two-dimensional grid and determined the best-fitting parameters in contrast to the primary transit's MCMC approach. The corresponding secondary eclipses and their fits are shown in Figs. \ref{fig:PP140_sec} and \ref{fig:PP127_sec}, respectively. For Wendelstein-2b, a small phase offset of 30\,min was detected, which is plausible since the largest possible offset for this system, based on an upper limit for the eccentricity $e$ of $0.057$ as measured by photometry, is $40$\,min. 
We estimated the planet's effective temperature by determining the central wavelength $\lambda$ of 3KK's H band wavelength response curve and the measured parameters for stellar effective temperature $T_\star$, radius $R_\star$ from Table \ref{tab:pp140-14711-stats} and planetary radius R$_{pl}$ from Table \ref{tab:pp140-14711-stats}. We measured the out-of-eclipse and eclipse fluxes $F'$ and $F$, respectively, in order to determine the flux ratio using Planck's law for $B_\lambda$: 
\begin{equation}
  \frac{F'}{F} = \frac{B_\lambda(\lambda, T_\star) \cdot R_\star^2 + B_\lambda(\lambda, T_{pl}) \cdot R_{pl}^2 }{B_\lambda(\lambda, T_\star) \cdot R_\star^2},
\end{equation}
which after a few transformations leads to:
\begin{equation}
  T_{pl} = \frac{h c}{k_B \lambda} \cdot \left( \ln \left( 1 + \frac{e^{\frac{h c}{k_B T_\star \lambda}}-1}{(\frac{F}{F'} -1) \cdot \frac{R_\star^2}{R_{pl}^2} \cdot } \right)\right)^{-1}.
\end{equation}
We combined the spectroscopic, SED-fitting based and GAIA-based effective temperatures from Table \ref{tab:pp140-14711-stats} for $T_\star$. As for the uncertainty of $T_{pl}$, we determined it by measuring the standard deviation of the residuals both during and outside of the secondary eclipse, combining them with the uncertainties for $T_\star$ and $R_\star$. That way, both the intrinsic variability of the light curve and uncertainties of the stellar parameters were taken into account.
However, the photometric noise and the sub-optimal phase coverage means that we cannot announce a safe detection of the secondary eclipses. Instead, we give a $1\sigma$ and $3\sigma$ upper limit for the likely effective temperature of the companion, which still rules out any stellar companions. We were also unable to detect secondary eclipses in the g'r'i'z'-band photometry. 
The results are shown in Tables \ref{tab:pp140_fitpara} and \ref{tab:pp127_fitpara}, respectively.

The challenge of these objects lies in their extremely faint brightness and our subsequent use of the H band where thermal emission contrast is even more shallow than in Ks. While one might conclude to use larger-aperture telescopes for more photons like the Wide-field InfraRed Camera (WIRcam) \citep{2004SPIE.5492..978P} on the Canada-France-Hawaii Telescope (CFHT) on Hawaii, it would be of limited use due to the high sky background in Ks. Only space missions or spectroscopic instruments such as OCTOCAM \citep{10.1117/12.2231862} on the Gran Telescopio Canarias (GTC) would be able to improve the contrast and this technique is therefore better suited for less faint stars of at least 12\,mag in Ks. 
\section{Discussion}
Multiband photometry is an efficient way of extracting additional information out of transit measurements, even more so when the different bands are recorded simultaneously since cross-correlation for improved systematics corrections is possible. Other groups \citep{2014A&A...567A..14T,2016ApJ...819...27F,2016MNRAS.456..990C} are also active in this field and have paved the way. However, 3KK's near-infrared capabilities allow for a more thorough validation than \citet{2014A&A...567A..14T} performed, while simultaneously measuring the system parameters similar to \citet{2016ApJ...819...27F,2016MNRAS.456..990C} with its optical channels. 

We followed up exoplanet candidates identified by the Pan-Planets survey. The faint nature of these targets and their low effective temperature present challenges for traditional high-resolution spectroscopic follow-up. It takes about 10\,h of 8\,m-class time and a dedicated instrument that is optimized for NIR for our planets to get confirmed via the traditional RV path, something that is infeasible for larger surveys or less massive planets. TESS \citep{2015JATIS...1a4003R} is focused on brighter targets in the magnitude range of $4\,$mag $< I_C < 13\,$mag, however, fainter sources can be extracted in the full frame images. We demonstrate that this concept can be used for such faint targets on 2\,m-class telescopes but the reached photometric accuracy is by no means final. Guiding and the installation of beam-shaping diffusers \citep{2017ApJ...848....9S} can significantly improve performance and allow ground-based telescopes to photometrically rival space telescopes. While the NIR channel of 3KK allows for a more thorough analysis of candidates, this technique can also be adapted for the optical infrared to give upper limits on the secondary eclipse depth. 

\begin{figure}
  \centering
  \includegraphics[width=\linewidth]{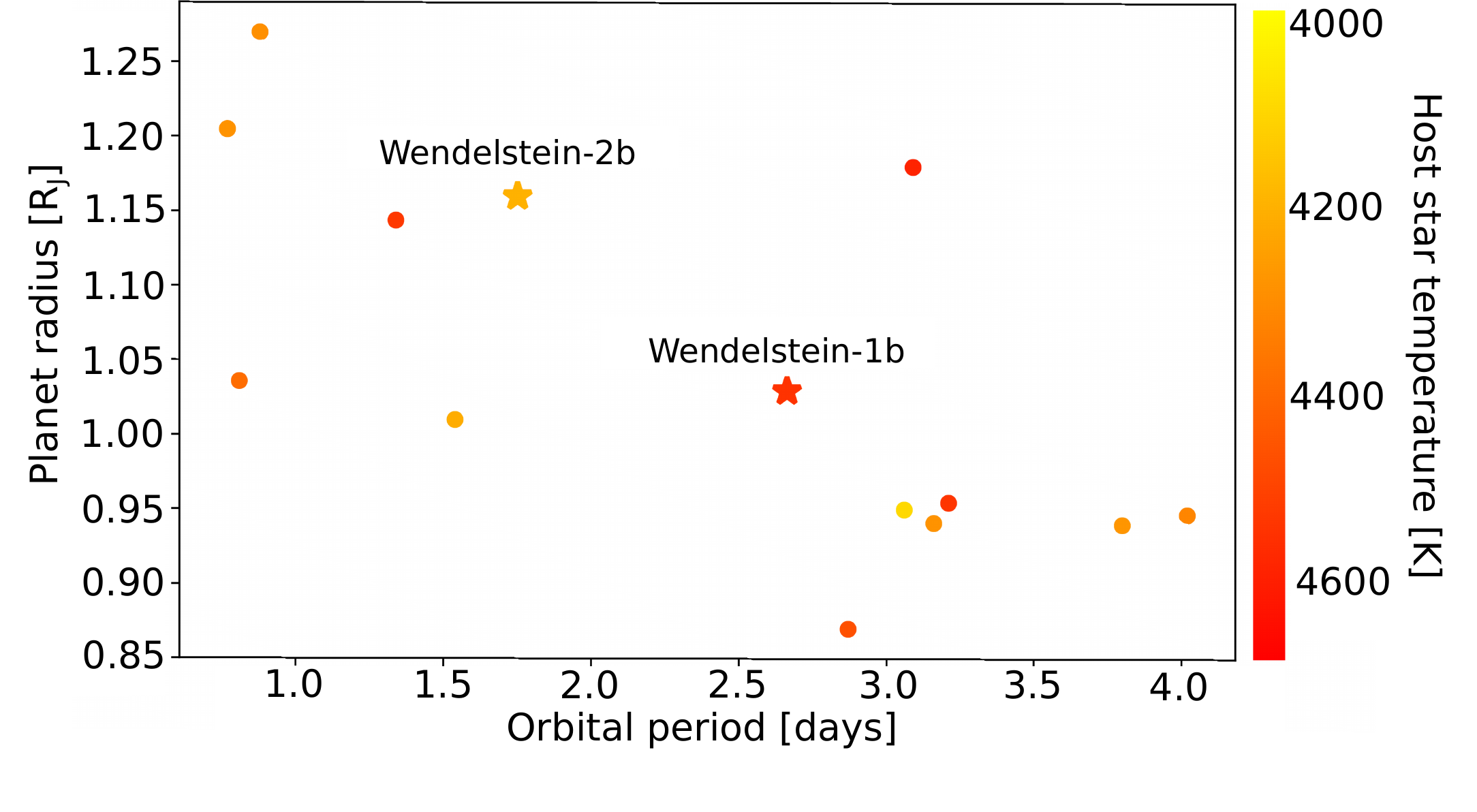}
  \caption{Orbital period as a function of the planetary radius for previously discovered similar exoplanets, color-coded for the effective temperatures of their host stars. Only host stars similar to Wendelstein-1b and 2b were selected for the parameter space, i.e., 4000\,K < $T_{\rm eff}$ < 4700\,K and 3.5 < log g < 5.0.}
  \label{fig:plot_distribution}
\end{figure}

The planets presented in this work are of scientific interest due to the rare combination of their large size, small semi-major axis and the cool nature of their host stars. 
While this combination appears to be ideal for detection, only a dozen planet systems of comparable (4000\,K < $T_{\rm eff}$ < 4700\,K, 3.5 < log g < 5.0) host star temperatures have been discovered to date, which are shown in Fig. \ref{fig:plot_distribution}. 
In general, there seemingly is a drop-off in the occurrence rate of hot Jupiters in the late K to M dwarf region. A possible explanation for this might be that protoplanetary disks of cool stars usually do not hold enough material for Jovian planets to form. It is not trivial to define a cut-off where enough material is available \citep{2012A&A...541A..97M,2010PASP..122..905J}. Previous efforts, either conducted using the radial velocity method (\cite{2007ApJ...670..833J}, \cite{2013A&A...549A.109B}) or transit method \citep{2014ApJ...791...10M,2015ApJ...807...45D,2013A&A...560A..92Z,2013MNRAS.433..889K}, concluded a best estimate of less than 1\% occurrence rate for these planets around M-dwarfs. These are just upper limits and further data is required to constrain this number further. 

We attempted to address this issue by determining an occurrence rate for hot Jupiters around cool host stars via the Pan-Planets survey \citep{2009A&A...494..707K,2016A&A...587A..49O} via observing a sufficiently large sample size of about 60000 M dwarfs. However, in the end the survey was unsuccessful in that regard due to its higher than expected photometric noise and subsequently less planet candidate detections than expected. All hot Jupiter candidates around M dwarfs were ruled out by the multiband observations. In \citet{2016A&A...587A..49O}, we estimated an upper limit for the M-dwarf hot Jupiter occurrence rate in case of a null result of 0.34\% with a 95\% confidence interval. 
The host stars of the planets in this work are categorized as late K-dwarfs, which still makes them quite rare with only a few other \citep{2012AJ....143..111J,2015AJ....149..166H,2013A&A...551A..80T} similar systems having been discovered so far.
\begin{acknowledgements}
\\The Pan-STARRS1 Surveys (PS1) have been made possible through contributions of the Institute for Astronomy, the University of Hawaii, the Pan-STARRS Project Office, the Max-Planck Society and its participating institutes, the Max Planck Institute for Astronomy, Heidelberg and the Max Planck Institute for Extraterrestrial Physics, Garching, The Johns Hopkins University, Durham University, the University of Edinburgh, Queen's University Belfast, the Harvard-Smithsonian Center for Astrophysics, the Las Cumbres Observatory Global Telescope Network Incorporated, the National Central University of Taiwan, the Space Telescope Science Institute, the National Aeronautics and Space Administration under Grant No. NNX08AR22G issued through the Planetary Science Division of the NASA Science Mission Directorate, the National Science Foundation under Grant No. AST-1238877, the University of Maryland, and Eotvos Lorand University (ELTE).
\\This paper contains data obtained with the 2.1m Fraunhofer
Telescope of the Wendelstein observatory of the Ludwig-Maximilians 
University Munich.
\\We thank the staff of the Wendelstein observatory for technical help and strong support, including observing targets for us, during the data acquisition.
\\This publication makes use of data products from the Two Micron All Sky Survey, which is a joint project of the University of Massachusetts and the Infrared Processing and Analysis Center/California Institute of Technology, funded by the National Aeronautics and Space Administration and the National Science Foundation.
\\This paper includes data taken at The McDonald Observatory of The University of Texas at Austin.
\\The Hobby-Eberly Telescope (HET) is a joint project of the University of Texas at Austin, the Pennsylvania State University, Stanford University, Ludwig-Maximilians-Universität München, and Georg-August-Universität Göttingen. The HET is named in honor of its principal benefactors, William P. Hobby and Robert E. Eberly. 
\\These results are based on observations obtained with the Habitable-zone Planet Finder Spectrograph on the Hobby-Eberly Telescope. We thank the Resident astronomers and Telescope Operators at the HET for the skillful execution of our observations of our observations with the Habitable-zone Planet Finder (HPF).
\\This work was partially supported by funding from the Center for Exoplanets and Habitable Worlds. The Center for Exoplanets and Habitable Worlds is supported by the Pennsylvania State University, the Eberly College of Science, and the Pennsylvania Space Grant Consortium. This work was supported by NASA Headquarters under the NASA Earth and Space Science Fellowship Program through grants NNX16AO28H and 80NSSC18K1114. We acknowledge support from NSF grants AST-1006676, AST-1126413, AST-1310885, AST-1517592, AST-1310875, the NASA Astrobiology Institute (NAI; NNA09DA76A), and PSARC in our pursuit of precision radial velocities in the NIR. Computations for this research were performed on the Pennsylvania State University’s Institute for CyberScience Advanced CyberInfrastructure (ICS-ACI).
\\
\\The authors are honored to be permitted to conduct observations on Iolkam Du'ag (Kitt Peak), a mountain within the Tohono O'odham Nation with particular significance to the Tohono O'odham people.
\end{acknowledgements}
\bibpunct{(}{)}{;}{a}{}{,} 
\bibliographystyle{aa} 
\bibliography{bibliography} 

\end{document}